\documentclass[prd,preprintnumbers,nofootinbib,superscriptaddress]{revtex4}

\usepackage{graphicx}
\usepackage{amssymb,amsmath}
\usepackage{epstopdf}
\usepackage{hyperref}
\def\a{\alpha}		\def\b{\beta}		
\def\d{\delta}		\def\e{\epsilon}		
\def\g{\gamma}				
					
\def\m{\mu}		\def\n{\nu}			
					\def\r{\rho}
\def\s{\sigma}		
							
					\def\D{\Delta}
			
			\def\P{\Pi}
\def\Q{\Theta}		\def\S{\Sigma}

		\def\ch{{\cal H}}

\def\be{\begin{equation}}
\def\ee{\end{equation}}

\def\Aodd{\hat{A}^{-}}
\def\hodd{\hat{h}^{\times}}
\def\Aoddc{\hat{A}^{-  *}}
\def\hoddc{\hat{h}^{\times  *}}

\def\pv{\vec{p}}
\def\qv{\vec{q}} 

\def\dtf{f'}
\def\dttf{f''}

\def\dpk{{\Delta_{\vec{k}}}}

\def\half{{1 \over 2}}

\def\Aev{\hat{A}^{+}}
\def\hev{\hat{h}^{+}}
\def\rev{\hat{r}}

\def\H{\mathbf{H}}
\def\P{\mathbf{\Phi}}
\def\Mone{\mathbf{M}_1}
\def\Mtwo{\mathbf{M}_2}
\def\Q{\mathbf{Q}}

\def \ktwo {{\psi'_{\vec{k}}}}

\def \rh {\hat{\rho}_A}

\def\order{\mathcal{O}}

\begin{document}

\preprint{CALT 68-2768}

\title{Primordial Power Spectra from Anisotropic Inflation}

\author{Timothy R. Dulaney}
\email{dulaney@theory.caltech.edu}

\author{Moira I. Gresham}
\email{moira@theory.caltech.edu}

\affiliation{California Institute of Technology, Pasadena, CA 91125}

\date{\today}                                           
\begin{abstract}
We examine cosmological perturbations in a dynamical theory of inflation in which an Abelian gauge field couples directly to the inflaton, breaking conformal invariance.  When the coupling between the gauge field and the inflaton takes a specific form, inflation becomes anisotropic and anisotropy can persist throughout inflation, avoiding Wald's no-hair theorem.  After discussing scenarios in which anisotropy can persist during inflation, we calculate the dominant effects of a small persistent anisotropy on the primordial gravitational wave and curvature perturbation power spectra using the ``in-in'' formalism of perturbation theory.  We find that the primordial power spectra of cosmological perturbations gain significant direction dependence and that the fractional direction dependence of the tensor power spectrum is suppressed in comparison to that of the scalar power spectrum. 
\end{abstract}
\pacs{...}
\maketitle

\section{Introduction}

Inflation gives a compelling explanation of the flatness, homogeneity, and isotropy of our Universe on large scales. It also generically predicts a nearly scale-invariant spectrum of density perturbations, which is consistent with our observations of the cosmic microwave background (CMB) and of structure formation. Because of these successes, the inflationary paradigm has dominated very early Universe cosmology in recent years.

In this paper we focus on the prediction of isotropy from inflation. The no-hair theorem of inflation states, roughly speaking, that an initially expanding, homogeneous universe with positive cosmological constant, $\Lambda$, and matter satisfying the dominant energy condition will become indistinguishable from a universe with de Sitter geometry on a time scale of $\sqrt{3/\Lambda}$ \cite{Wald:1983ky}. Because of the no-hair theorem, isotropy is generally taken as a prediction of inflation.

But there could be ways around the no-hair theorem. For example, models with spacelike vector fields that get vacuum expectation values can lead to a preferred direction during inflation, evading the no-hair theorem because the vector field stress-energy tensor does not satisfy the dominant (or even the weak) energy condition \cite{Ackerman:2007nb}. However, such ``aether'' models have been shown to be unstable \cite{Dulaney:2008ph,Himmetoglu:2008hx,Carroll:2009em}.

Recently, another model has been shown to support a persistent anisotropy during inflation \cite{Watanabe:2009ct}. In this model, there is a nonminimal coupling between a $U(1)$ gauge field and the inflaton, essentially leading to a time-dependent $U(1)$ charge during inflation:
\begin{equation}\label{action}
S = \int d^4 x \sqrt{-g} \left[ {R \over 2 \kappa^2}  -{1 \over 2} (\partial_\mu \phi)(\partial^\m \phi) - V(\phi) -{f^2(\phi) \over 4} F_{\mu\nu}F^{\mu\nu}\right].
\end{equation}
Here, the $U(1)$ field strength, $F_{\mu \nu}$, may or may not be the electromagnetic field strength. When the coupling, $f(\phi)$, between the inflaton, $\phi$, and the $U(1)$ field takes a particular form and there exists a nonzero homogeneous $U(1)$ seed field, an anisotropy persists throughout inflation even though the space-time is undergoing nearly exponential expansion. More specifically, the ``electric'' field contributes non-negligible extra negative pressure in the direction in which it points, which causes space-time to expand more slowly in that direction.

The model avoids the no-hair theorem by having (1) expansion that is not purely exponential and (2) a coupling between the inflaton and other matter. The mechanism for evasion of the no-hair theorem shows up in our results in the following ways:  (A) all modifications to power spectra associated with the anisotropy go to zero when slow-roll parameters vanish and (B) isotropic dynamics is quickly restored if the inflaton-dependent coupling that breaks conformal invariance goes to a constant (as is the case at the end of inflation, when the inflaton field relaxes to the minimum of its potential). 

All of the standard energy conditions are satisfied in this model, which means it should not be plagued by stability issues as in aether models. The model does, however, suffer from the standard fine-tuning problems of single field inflation. Nevertheless, to our knowledge this model could be the first consistent model of inflation that evades the no-hair theorem and includes anisotropy at a significant level. It is therefore interesting to investigate whether the model is truly consistent and to investigate its potential astrophysical signatures.

To that end, in this paper we consider gauge-invariant cosmological perturbations in this anisotropic inflation model. We consider and discuss a model generalized from that of \cite{Watanabe:2009ct}, and extend their formula for the relation between the anisotropic expansion parameter and the slow-roll parameter to include arbitrary forms of the inflaton potential. We also present the dominant effect of the anisotropy on the power spectra of tensor, vector and scalar perturbation correlations at the end of inflation. 

Our main conclusions are:
\begin{itemize}
\item The power spectra for gravitational wave and curvature perturbations can develop dramatic direction dependence for very small values of the anisotropy parameter\footnote{The anisotropy parameter is basically the fractional difference between the rate of expansion in the preferred direction and that of a perpendicular direction.} if the parameter is nearly constant for a large period of inflation. 
\item The main cause of direction-dependence of the power spectra is a coupling between the $U(1)$ vector degrees of freedom to both tensor and scalar degrees of freedom through the anisotropic background. These interactions significantly affect the power spectra of modes after horizon crossing.
\item The ratio of the fractional direction-dependent change in the gravitational wave power spectrum over that of the curvature perturbation power spectrum is nearly equal to the tensor-to-scalar ratio. In particular, the curvature perturbation power spectrum has much stronger direction dependence than the gravitational wave power spectrum.
\item For a given scale, the tensor and scalar power in modes with wave vector perpendicular to the preferred direction is greater than the power in modes with wave vector parallel to the preferred direction.\footnote{\emph{I.e.}, the parameter $g_*$ (see equation \eqref{gstar}), as defined in \cite{Ackerman:2007nb}, that characterizes the direction-dependence of the power spectrum due to a preferred direction is \emph{negative}.} 
\item There is no indication that the anisotropic inflation model is unstable. (\emph{e.g.} There are no ghosts.) This should be unsurprising since the stress-energy tensor for matter in the model satisfies the dominant energy condition.  
\end{itemize}

Many have studied inflationary scenarios with actions similar to \eqref{action}, interpreting $F_{\mu \nu}$ as the standard model electromagnetic field strength, in the context of explaining the existence of large-scale magnetic fields in the Universe. 
Initially Parker \cite{Parker:1968mv} and then Turner and Widrow \cite{Turner:1987bw} showed that magnetic fields produced in an inflationary Universe are ``uninterestingly small'' (\emph{i.e.},~too small to possibly account for the observed large-scale magnetic fields in the Universe) unless the conformal invariance of the electromagnetic field is broken. The generation of seed magnetic fields starting from the action in \eqref{action} and a particular $f(\phi)$ was considered in \cite{Ratra:1991bn} and more recently in \cite{Demozzi:2009fu}.   Generic predictions for magnetic fields in a large class of models, of which the model we consider here is an example, were presented by Bamba, \emph{et.~al.}~\cite{Bamba:2008my}; the particular realization of the model we consider in this paper is what these authors refer to as the ``weak coupling case''.  Magnetogenesis, including the backreaction due to electromagnetic fields, in the inflationary scenario we consider here was considered in \cite{Kanno:2009ei}.  For a review of the generation of magnetic fields during inflation in a more general context see, for example, \cite{Grasso:2000wj}. 

More recently, the effect of vector fields during inflation has been studied in the context of their effects on the curvature perturbation power spectrum. A ``vector curvaton'' scenario, in which a vector field with time-varying mass and Maxwell-type kinetic coupling term contributes to the curvature power spectrum, was found in \cite{Dimopoulos:2009vu} to allow significant anisotropic contributions to the curvature spectrum and bispectrum if the vector field remains light until the end of inflation. A similar massless vector curvaton scenario was considered in  \cite{Yokoyama:2008xw} and again the possibility of significant anisotropic contributions was found.\footnote{Both studies employed the $\delta N$ formalism in calculating the curvature perturbation power spectra.} The anisotropic contribution of vector field perturbations to primordial curvature perturbation correlations in various inflationary scenarios was also considered in \cite{Lim:2004js,Koivisto:2008xf,Golovnev:2009ks,Dimopoulos:2008yv,ValenzuelaToledo:2009af,ValenzuelaToledo:2009nq}.
Perturbations of what correspond to our cross polarization gravitational wave degree of freedom were studied in \cite{Himmetoglu:2009mk}, but in a scenario in which a second scalar field, uncoupled to the $U(1)$ field and the scalar field that couples to the $U(1)$ field, causes a transition back to isotropic expansion before the end of inflation.   

This paper is organized as follows.  In section \ref{Model}, we introduce the model.  In section \ref {sec:setup}, we discuss our philosophy and methods for calculating and analyzing primordial perturbation spectra. Finally, in sections \ref {sec:odd_sector} and \ref {sec:even_sector} we calculate the primordial perturbation spectra and briefly discuss stability. We summarize our conclusions in section \ref {sec:conclusions}.

\section{Model and background solution}\label{Model}
We consider a space-time governed by the following action \cite{Watanabe:2009ct}:
\begin{equation}
S = \int d^4 x \sqrt{-g} \left[ {R \over 2 \kappa^2}  -{1 \over 2} (\partial_\mu \phi)(\partial^\m \phi) - V(\phi) -{f^2(\phi) \over 4} F_{\mu\nu}F^{\mu\nu}\right],
\end{equation}
where $g = \det(g_{\m\n})$, $R$ is the Ricci scalar, $\phi$ is the inflaton, and $F_{\m\n} = \partial_\mu A_\nu - \partial_\nu A_\mu$ is a $U(1)$ gauge field strength. For convenience, we'll refer to the $U(1)$ field as the ``electromagnetic'' (EM)  field, even though it need not be the standard model EM field. Here we've defined 
\be
\kappa^2 \equiv 8 \pi G  = 1 / M_\text{Planck}^2.
\ee

We assume that the background is homogeneous, and that there is a nonzero homogeneous electric field.\footnote{At least we assume that the ``electric'' field was aligned in our causal patch. We will not consider the effects of regions with differing directions of alignment of the electric field.}  We orient coordinates such that $F_{i j} = F_{\eta y} = F_{\eta z} = 0$ and $F_{\eta x \neq 0}$.  One could just as easily have chosen to consider a homogeneous magnetic field.  This choice does not change the form of the background stress tensor, and we expect the results of this paper to apply in the magnetic field case as well.  However, allowing for both electric and magnetic fields of arbitrary relative alignment is beyond the scope of this paper.  

The background space-time is Bianchi I and the metric can be written in the following form by appropriate choice of coordinate axes:\footnote{The form is chosen so that the spatial metric has unit determinant (and therefore scaling or translating $\beta(\eta)$ does not affect the spatial volume element).} 
\begin{equation} \label{Bianchi1}
ds^2 = a(\eta)^2 \left(- d\eta^2 + \gamma_{ij}(\eta) dx^i dx^j \right),
\end{equation}
where\footnote{An equivalent ansatz would have been: $ds^2 = -dt^2 + a_{\parallel}(t)^2 dx^2 + a_{\perp}(t)^2 (dy^2 + dz^2).$}
\be
\gamma_{xx} = e^{-4\beta(\eta)},~~\gamma_{yy}=\gamma_{zz} = e^{2\beta(\eta)}~~\text{and} \qquad \gamma_{ij} = 0~~\text{ for all }~~i \neq j.  
\ee
Since $g$ is independent of $\beta$, the scale factor, $a$, completely characterizes the space-time volume. For convenience we define $\alpha$ to be the logarithm of the scale factor, so
\be
 a=e^\alpha.
\ee
In parametrizing the metric, we've used the conventions of \cite{Pereira:2007yy}.  
The solution to the background electromagnetic field equation of motion is then  \cite{Watanabe:2009ct},
\begin{equation}
F_{\eta x} = p_A {e^{-4 \beta(\eta)} \over f^2(\bar{\phi})},
\end{equation}
where $p_A$ is an integration constant of mass dimension two and a prime indicates a derivative with respect to conformal time $\eta$.  In these coordinates, Einstein's equations take the form  \cite{Watanabe:2009ct}
\begin{eqnarray}
\alpha'^2 &=& \beta'^2 + {\kappa^2 \over 3} \left[{ \phi'^2 \over 2} + e^{2\alpha} V(\bar{\phi}) + {p_A^2 e^{-2\alpha - 4\beta} \over 2 f^2(\bar{\phi})}\right], \label{Einstein00} \\
\alpha'' &=& - 2 \alpha'^2 + \kappa^2 e^{2\alpha} V(\bar{\phi}) + {p_A^2  \kappa^2 e^{-2\alpha - 4\beta} \over 6 f^2(\bar{\phi})}, \label{TrEinstein} \\
\beta'' &=& - 2 \alpha' \beta' + {p_A^2 \kappa^2  e^{-2\alpha - 4\beta} \over 3 f^2(\bar{\phi})}. \label{AnisoEinstein}
\end{eqnarray}
Given Einstein's equations above, the equation of motion for $\phi$ is redundant.\footnote{Recall that Einstein's equations and the matter field equations are related through the conservation equation, $\nabla_\m T^\m_\n = 0$, where $T^\m_\n$ is the matter stress-energy tensor.}

It was shown that inflation can occur for suitable initial conditions such that the Universe is initially expanding, and that the energy density of the vector field will remain almost constant with respect to the inflaton energy density if $f(\phi) \propto e^{-2\alpha}$ \cite{Watanabe:2009ct}. (Recall that if there's no inflaton-electromagnetic coupling, the ratio of electromagnetic energy density to inflaton energy density decays as $a^{-4}$.) Let us briefly show how this can occur.

If expansion is nearly exponential (in cosmic time), then the ``slow-roll'' parameters, 
\begin{align}
\epsilon  & \equiv - {\partial_t H \over H^2} = {\alpha'^2 - \alpha''  \over \alpha'^2}  \\
\delta & \equiv { \partial_t^2 H \over 2 H \partial_t H}
\end{align}  
are very small compared to one and as usual, $H \equiv {\partial_t a \over a}$.\footnote{Note that 
\be
{\epsilon ' \over \alpha'}  = 2 \epsilon (\epsilon + \delta).
\ee}
Higher derivatives of $H$ must, of course, also be small if expansion is nearly exponential.

The field equations \eqref{Einstein00}, \eqref{TrEinstein} and \eqref{AnisoEinstein}, can be cast in the following form:
\begin{align}
\rh &\equiv {\kappa^2 p_A^2 e^{-4\beta} \over 2 a^2 f^2(\bar{\phi}) \alpha'^2} = {3 \over 2}\left( 3 \Sigma - \epsilon \Sigma + {\Sigma' \over \alpha'}\right)  \label{PotentialEqn2} \\
\hat{\rho}_\phi &\equiv {a^2 \kappa^2 V(\bar{\phi}) \over \alpha'^2} = 3 - \epsilon - {3 \over 2} \Sigma + {\epsilon \over 2} \Sigma - {\Sigma' \over 2\alpha'} = 3 - \epsilon - {1 \over 3} \rh \label{PotentialEqn1} \\
{\kappa^2 \bar{\phi}'^2 \over \alpha'^2} &= 2 \epsilon - 6 \Sigma + 2 \epsilon \Sigma - 6 \Sigma^2 - 2 {\Sigma' \over \alpha'} = 2 \epsilon - {4 \over 3} \rh - 6 \Sigma^2 \label{eq:phiprime} \\
\text{where}~~~\Sigma &\equiv \b' /\a'.
\end{align}
The quantities $\hat{\rho}_\phi$ and $\rh$ are dimensionless energy densities, normalized by the Hubble scale squared times the Planck mass squared.

In standard single field inflation with an inflaton potential $V$, for example, one finds from the field equations that ${\kappa \phi ' \over \alpha '} \sim \sqrt{\epsilon}$, so that if expansion is nearly exponential, then the inflaton must be slowly rolling. Taking derivatives of the above equations in the isotropic case, one can find expressions for derivatives of $V$ in terms of slow-roll parameters---thus yielding requirements of a potential that can give rise to inflation.

From \eqref {PotentialEqn1} and \eqref {PotentialEqn2} one finds
\begin{align}
{\hat{\rho}_\phi '  \over  \hat{\rho}_\phi \alpha'} &= {\partial_\phi V \over \kappa V}{\kappa \bar{\phi}' \over \alpha'} + 2 \epsilon   = {   - {\epsilon' \over \alpha'} - {1 \over 3}{\rh' \over \alpha'}  \over 3 - \epsilon - {1 \over 3 } \rh} \label{PotentialEqn1prime} \\
{\rh' \over \rh \alpha'}&= -4 - 2 {\partial_\phi f \over \kappa f}{\kappa \bar{\phi}' \over \alpha'} + 2 \epsilon - 4 \Sigma = { 2 {\Sigma' \over \alpha'}  + \ldots  \over 3 \Sigma - \epsilon \Sigma + {\Sigma' \over \alpha'} } \label{PotentialEqn2prime} 
\end{align} where $\ldots \sim \mathcal{O}(\Sigma {\epsilon' \over \alpha'}, \epsilon {\Sigma' \over \alpha'}, {\Sigma'' \over \alpha'^2})$.  

We can glean a fair bit of information from equations \eqref {PotentialEqn2} - \eqref {PotentialEqn2prime} without much effort. First, what if expansion were purely exponential so that $\delta = \epsilon = 0$? From \eqref {eq:phiprime} we can immediately see that $\rh$ and $\Sigma$ had better  then also be zero based simply on the fact that ${\kappa^2 \bar{\phi}'^2 \over \alpha'^2} $, $\rh$, and $\Sigma^2$ are positive. This could be seen as confirmation of the no-hair theorem; anisotropy  can exist only if expansion is \emph{not} purely exponential.\footnote{%
A more direct confirmation of the no-hair theorem comes from supposing $\phi' = 0$ (and, for simplicity, $\epsilon << 1$) so that $V(\phi)$ functions as a cosmological constant. Then from \eqref{eq:phiprime} and \eqref {PotentialEqn1}
\be
{d \log \rh \over d t }  \approx - 4 {d \over d t}\alpha \approx -4 \kappa \sqrt{V(\phi) \over 3}.
\ee
So $\rh$, and thus by \eqref {PotentialEqn2} also $\epsilon$ and $\Sigma$,  go to zero on the time scale promised by the no-hair theorem.
} Similarly, if $\epsilon$ is small, then $\rh$ and $\Sigma$ had also better be small.  In particular, even in small field models of inflation where typically $\epsilon \ll \delta \ll 1$, the anisotropy parameters $\Sigma$ and $\rh$ must be order $\epsilon$ or smaller. Second, from \eqref{PotentialEqn1prime} we see that $\hat{\rho}_\phi$ is nearly constant with respect to the Hubble parameter if $\epsilon$ and $\Sigma$ are small.  Also from \eqref {PotentialEqn1prime}  we see that 
\be
 {\partial_\phi V \over \kappa V}{\kappa \bar{\phi}' \over \alpha'} = -  2 \epsilon + \mathcal{O}(\epsilon ' / \alpha ' ) + \ldots
\ee
Third, from \eqref{PotentialEqn2prime} , if $\epsilon$ and $\Sigma$ are small, we see that $\rh$ decreases rapidly with respect to the Hubble parameter \emph{unless} 
\be
{f ' \over f \alpha' } \lesssim - 2
\ee
or equivalently unless
\be
{\partial_\phi f \over \kappa f} \lesssim - 2 /  \left( {\kappa \bar{\phi}' \over \alpha'} \right) .
\ee

Now since 
\be 
\left( {\partial_\phi V \over \kappa V} \right)^{-1} \sim \pm \sqrt{ 1/2 \epsilon } \sqrt{1 - 3 \Sigma/\epsilon + \ldots}  \sim - \left( \kappa \bar{\phi}' \over \alpha' \right)^{-1}
\ee
a ready choice for the coupling function, $f$, if one wants the energy density of the electromagnetic field (and thus the anisotropy) not to decay rapidly with respect to the inflaton energy density, is thus
\be\label{eq:fofphi}
f(\phi) = \exp \left\lbrace 2 c \kappa \int  \left( {\partial_\phi V \over \kappa V} \right)^{-1} d \phi \right\rbrace
\ee
where $c$ is an order one constant. This is the coupling function motivated and examined in \cite{Watanabe:2009ct}. Let's suppose the coupling function is of this exact form, so
\begin{align}
{\rh' \over \rh \alpha'} &= -4 - 4 c \left({\kappa \bar{\phi}' \over \alpha'}\right)^2 \left( {\partial_\phi V \over \kappa V} {\kappa \bar{\phi}' \over \alpha'} \right)^{-1} + 2 \epsilon - 4 \Sigma \label{eq:rhoAprime1} \\
	 &= -4 - 4 c (2 \epsilon - 6 \Sigma + \ldots)(-  2 \epsilon + \mathcal{O}(\epsilon ' / \alpha ' ) + \ldots)^{-1} + 2 \epsilon - 4 \Sigma \label {eq:rhoAprime2}\\
	  &= (c-1)4 - 4 (3 c )  {\Sigma \over \epsilon} + \ldots \label {eq:rhoAprime3}
\end{align}
Suppose initially that $\Sigma \ll \epsilon$. If $c < 1$ then $\rh$ decreases along with $\Sigma$ as long as $\epsilon$ is small. Anisotropy is wiped out (albeit much more slowly than in the case where $f(\phi) = 1$). If $c > 1$, then $\rh$ initially increases, as does $\Sigma$ (see \eqref {PotentialEqn2}). The derivative of the electromagnetic field energy density will thus approach zero, ${\rh' \over \rh \alpha'} \longrightarrow 0$, and so $\rh$ and $\Sigma$ will become nearly constant for a time. If $\Sigma$ is initially greater than ${(c - 1) \over 3 c} \epsilon $, then $\rh$ and $\Sigma$ will initially decrease, $\phi$ will climb its potential, and then it will fall back down (slowly) after $\Sigma$ has approached a constant \cite{Watanabe:2009ct}.

From \eqref{PotentialEqn2} one can see that if $\Sigma$ is approximately constant then $\Sigma$ must be positive. So when the space-time undergoes anisotropic expansion in this model (and $\Sigma$ is nearly constant) the preferred direction expands more slowly than the perpendicular directions.

When \eqref{eq:fofphi} holds, we can find an expression for $\Sigma$ in terms of the slow-roll parameter during the period in which it is nearly constant. Assuming
\be
\mathcal{O}(\epsilon) \approx \mathcal{O}(\delta) \qquad c - 1 >  \mathcal{O}(\epsilon) \qquad \Sigma \lesssim \mathcal{O}(\epsilon), \qquad { \Sigma '  \over \alpha' } \lesssim \mathcal{O}(\epsilon  {\Sigma} ), \qquad \left( { \Sigma ' \over \alpha' }\right)' / \alpha'  \lesssim \mathcal{O}(\epsilon^2  {\Sigma} )
\ee
we can set the two different expressions for $\partial_\phi V / V$ derived from equations \eqref {PotentialEqn1prime} and \eqref {PotentialEqn2prime}  equal to each other. Using this method we find
\begin{equation} \label{eq:betaprime}
\Sigma \equiv {\beta' \over \alpha'} = {c-1 \over 3 c} \epsilon + {1+c-4c^2 \over 18c^2} \epsilon^2 + {1-2c-4c^2 \over 18c^2} \epsilon \delta + \ldots \qquad \text{assuming $c - 1 >  \mathcal{O}(\epsilon) $}.
\end{equation}
The authors of \cite{Watanabe:2009ct} derived this expression to first order in $\epsilon$ for the particular potential $V = \half m^2 \phi^2$ and argued that $\Sigma$ generically tracks the slow-roll parameter for general potentials. We find that the expression \eqref {eq:betaprime}  actually holds for any potential $V$ in a slow-roll regime ($\epsilon, \delta \ll 1$). 

As $c \rightarrow 1$, the story is a bit different. For example, if $c = 1$, looking back to equations \eqref{eq:rhoAprime1} - \eqref{eq:rhoAprime3} one finds that $\rh$, if it is initially greater in magnitude than $\mathcal{O}(\epsilon^2)$, decreases until it's on the order of $\epsilon^2$, and then stays nearly constant. From numerical studies it appears that if  $\rh$ is initially much greater in magnitude than $\mathcal{O}(\epsilon^2)$, then it will rapidly settle to a value much smaller than $\mathcal{O}(\epsilon^2)$. If the magnitude of  $\rh$  is initially on the order of $\epsilon^2$ or less, then it will stay very nearly constant until the end of inflation. An example with $c = 1$ is provided in Fig. \ref{fig:sigma_plot}.
{
\begin{figure}
 \centering
\includegraphics[width=0.5\textwidth]{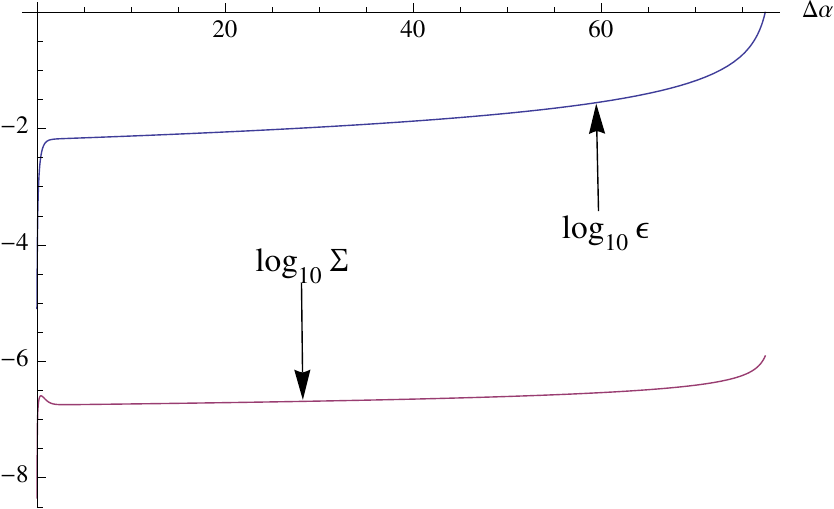}
\caption{Log plot of $\Sigma$ and $\epsilon$ as a function of $e$-foldings ($\Delta \alpha = \alpha - \alpha_0$) during inflation. The plot was generated with the potential $V = \half m^2 \phi^2$ and coupling function $f(\phi) = \exp\left[{\kappa^2 \phi^2 \over 2}\right]$.  The initial conditions were $\phi_0 = 17.5/\kappa$, $\phi_0' = 0$, $\alpha_0 = -75$, $\beta_0 = 0$ and $\beta_0' = 0$.  The constants $m$ and $p_A$ were chosen so that initially $\rho_A/\rho_\phi \approx 10^{-6}$. Notice that $\Sigma$ very quickly settles to a value that is somewhat smaller than the square of the slow-roll parameter $\epsilon$. }\label {fig:sigma_plot}
\end{figure}
}

The trick of this model is to choose $f(\phi)$, given $V(\phi)$, such that the electromagnetic field energy density does not decay rapidly with respect to the inflaton energy density during inflation. We saw above that a choice guaranteed to work is 
\be
f(\phi) \approx \exp \left\lbrace 2  \kappa \int  \left( {\partial_\phi V \over \kappa V} \right)^{-1} d \phi \right\rbrace.
\ee
For example, if $V(\phi) \propto \phi^n$, then $f(\phi) \approx \exp [ \kappa^2 \phi^2/n]$. What if we were to choose instead, say, $f(\phi) \approx \exp [ \lambda \kappa \phi]$? Then we would have
\be
{\rh' \over \rh \alpha'} = -4 - 2 \lambda {\kappa \bar{\phi}' \over \alpha'} + 2 \epsilon - 4 \Sigma. 
\ee
If $\lambda$ is order one, then the anisotropy will rapidly decay. However, if $\lambda$ were large enough in magnitude then the anisotropy could persist for a good portion of inflation.

In our analysis, we will use only the background equations of motion, leaving $f(\phi)$ and $V(\phi)$ generic. We will then be interested in scenarios in which anisotropy can persist over several $e$-folds---scenarios in which ${f' \over f \alpha'} = -2 + \mathcal{O}(\epsilon)$ and where $\rh \approx 9 \Sigma / 2 $ is approximately constant. We saw that consistency of the background equations and a slow-roll scenario dictates that $\rh$ must be order $\epsilon$ or smaller. We also discussed specific examples of functions, $f(\phi)$, that can lead to such scenarios (assuming, otherwise, a slow-roll scenario, $\epsilon, \delta \ll 1$). In order to calculate primordial power spectra, we will use the ``in-in'' formalism of perturbation theory, assuming\\
\begin{itemize}
\item $\epsilon \ll 1$, $\delta \ll 1$\\
\item $\rh \approx 9 \Sigma / 2  \lesssim \mathcal{O}(\epsilon)$\\
\item $\rh' / (\rh \alpha') \lesssim \mathcal{O}(\epsilon)$.
\end{itemize}


\section{Perturbations: setup and strategy}\label{sec:setup}

Our goal is to examine whether the background described in the previous section (slightly generalized from the space-time of \cite{Watanabe:2009ct}) is perturbatively stable, and to examine its signature at the level of primordial perturbation spectra.   

We have calculated the quadratic action for dynamical modes in terms of the gauge-invariant variables defined in appendix \ref {app:parameterization}. We calculated the action to quadratic order in perturbations starting with the form of the second order Einstein-Hilbert action given in appendix \ref {ap:einsteinHilbert}, and a similar expression for the quadratic-order matter action. We worked in Newtonian gauge and used a differential geometry package in \emph{Mathematica} to massage the quadratic action into the (relatively) simple, manifestly gauge-invariant form presented in sections \ref {sec:odd_sector} and \ref {sec:even_sector}. 

Regarding perturbative stability of the background, we find that there are no ghosts (fields with wrong-sign kinetic terms), and no other indication of instability at the quadratic level. Here, we take ``perturbative stability'' to mean that dimensionless combinations of fields assumed to be much less than one in the perturbative expansion of the action remain small.  We find that such small quantities do indeed stay small.

In the remainder of this section we describe how we set up the calculation and analysis of perturbation spectra; we describe the physical scenario, the expression for expectation values in the ``in-in'' formalism, the definitions for the relevant degrees of freedom, and, finally, the current bound on a preferred direction during inflation. In sections \ref {sec:odd_sector} and \ref {sec:even_sector} we calculate power spectra and briefly discuss stability.

\subsection{Physical scenario}

Perturbations from inflation are usually assumed to be generated in the following way \cite{Mukhanov:1990me}:
\begin{itemize}
\item Quantum mechanical perturbative modes are in their ground state throughout inflation. So the vacuum expectation value of individual modes is zero, though the variance is generally nonzero.
\item The normalization of the ground states is such that when the modes are well within the horizon, the canonically normalized\footnote{In conformal time, the kinetic term for a canonically normalized field, $\phi$, in the quadratic action takes the form ${1 \over 2} \phi'^2 $.} fields, $\phi$, obey a simple harmonic oscillator equation and satisfy the canonical commutation relations.\footnote{Specifically,
\begin{equation}
  [  \partial_\eta \phi(\eta, \vec{x}), \phi(\eta, \vec{y} ) ] = - i \hbar \delta^3(\vec{x} - \vec{y}) 
\end{equation} where $\eta$ is conformal time.}
\item As modes cross the horizon, their correlations are ``frozen in'' and translate into classical perturbations that lead to, for example, density perturbations that seed the formation of structure in the Universe and lead to temperature anisotropies of the cosmic microwave background radiation. 
\end{itemize}

We shall assume the same, with one complication. We assume the quantity, $\Sigma \equiv \beta ' / \alpha' $, which characterizes the deviation from isotropy, is nonzero so that expansion of the background space-time is slightly anisotropic, and modes that corresponded to scalar, vector, and tensor degrees of freedom in the isotropic background are now coupled. (Several scenarios in which this can occur were discussed in section \ref{Model}.) Because of the coupling of modes, the amplitudes of tensor, vector, and scalar perturbations are not separately conserved outside the horizon.  As the inflaton decays at the end of inflation, the dynamics becomes isotropic again, and tensor, scalar, and vector modes decouple. At this point, superhorizon perturbations should be frozen in. We are therefore interested in the correlations of perturbations at the end of inflation. Especially if the $U(1)$ field in our model were interpreted as the electromagnetic field, the details of the reheating process at the end of inflation could also be important in calculating the direction dependence of CMB power spectra. In this paper, however, we will only examine the effects of the gauge field on curvature and gravitational wave power spectra until just before reheating.

\subsection{Correlations using ``in-in'' formalism}

Because in the context of cosmological perturbations as described above we know only the quantum ``in'' states and we're interested in expectation values evaluated at a particular time, we use the ``in-in'' formalism of perturbation theory (see \emph{e.g.}~\cite{Weinberg:2005vy}).  We separate our Hamiltonian into a free portion $H_0$ and an interacting portion $H_I$.  The interaction-picture (free) fields' evolution is determined by the free Hamiltonian.  The expectation value for a general operator $X$ at (conformal) time $\eta$ can be written as
\begin{equation} \label{perturbedpower}
\left< X(\eta) \right> = \left< X^I(\eta) \right> + i \int^\eta d \eta' \left< [H_I(\eta'), X^I (\eta)] \right>  + (i)^2\int^\eta d \eta'  \int^{\eta'} d \eta''  \left< [H_I({\eta'}) ,[H_I(\eta''), X^I (\eta)] ] \right>+ \ldots
\end{equation}
where the ellipsis denotes terms with more powers of $H_I$ and where $X^I$ is the interaction-picture operator.

It should be noted that corrections of quadratic (or higher) order in the interaction Hamiltonian can lead to ambiguities when the details of the contour integration are not carefully considered \cite{Adshead:2008gk}.   We will work only to linear order in $H_I$, and therefore we need not worry about such ambiguities.

\subsection{Decomposition of perturbations}\label{subsec:decomp}

Since the background space-time is homogeneous, we decompose our perturbations into Fourier modes
\be
\delta(x^i,\eta) = \int {d^3 k \over (2\pi)^{3}}~ e^{ i k_j x^j} \delta (k_i, \eta).
\ee

We analyze perturbations about an anisotropic background. Since the background is anisotropic and thus there is no $SO(3)$ symmetry, perturbations cannot be decomposed into spin-0, spin-1, and spin-2 degrees of freedom and analyzed separately. We instead decompose gauge-invariant perturbations according to their transformation properties in the isotropic limit. (See appendix \ref{app:parameterization}.)

There are five dynamical degrees of freedom in our model, corresponding to 
\begin{itemize}
\item one scalar degree of freedom, $r$ (spin-0 in isotropic limit),\footnote{See \eqref{eq:r} in appendix \ref{app:parameterization}.}
\item two electromagnetic vector degrees of freedom, $\delta A^+$ and $\delta A^-$ (spin-1 in isotropic limit),
\item and two metric tensor degrees of freedom, $E^+$ and $E^\times$ (spin-2 in isotropic limit).
\end{itemize}

In order to analyze the relevant dynamical perturbative degrees of freedom in our scenario, we derived the quadratic action in terms of the gauge-invariant variables of appendix \ref{app:parameterization}. Then we eliminated the nondynamical degrees of freedom by using constraint equations derived from the action.  Finally, we canonically normalized the degrees of freedom that correspond to the dynamical ``free'' fields in the limit as $\beta' / \alpha' \longrightarrow 0$.  Within the ``in-in'' formalism of perturbation theory, we take the interaction-picture fields to be those governed by the dynamics in the $\beta' / \alpha' = 0$ limit. 

The quadratic action separates into two uncoupled pieces according to a residual symmetry under parity transformations. (See appendix \ref{app:parameterization}.) The ``odd'' sector has two degrees of freedom, $E^\times$ and $\delta A^{-}$. The ``even'' sector has three degrees of freedom, $E^+$, $\delta A^+$, and $r$.
The fields $E^+, E^\times$, and $r$ correspond to fields that are conserved outside the horizon during isotropic inflation.  Here $r$ is a Mukhanov-Sasaki variable, equal to minus the curvature perturbation, $- \zeta$, as defined in, \emph{e.g.}~\cite{Dodelson:2003ft}, in a gauge with spatially flat slicing. We will therefore refer to $r$ as the curvature perturbation.

\subsection{Canonically normalized variables}

The canonically normalized fields in each sector (``even" and ``odd'', respectively) are given by
\begin{equation}\label{eq:fieldDefs}
\begin{array}{r c l}
 \vspace{1 ex}
 \hat{A}^+ &=&  f(\bar{\phi})\, \delta A^+  \\
 \vspace{1 ex}
 \hat{h}^+ &=& {a(\eta) }\, E^{+} / \kappa  \\
 \hat{r} &=& z(\eta) \, r 
\end{array}
\qquad
\text{and}
\qquad
\begin{array}{r c l}
 \vspace{1 ex}
 \hat{A}^- &=& { f(\bar{\phi}) \, \delta A^- }\\
 \hat{h}^\times &=& { a(\eta) } \, E^{\times}  / \kappa
\end{array}
\end{equation}
where
\begin{equation}\label{eq:z}
z(\eta) \equiv a(\eta){\bar{\phi}' \over \alpha'}.  
\end{equation}

The fields on the right-hand sides of equations \eqref {eq:fieldDefs} are defined in appendix \ref {app:parameterization}. As mentioned above, in the isotropic limit $E^+, E^-$ and $r$ are conserved outside the horizon. The other  important fact about the fields above is that the perturbative expansion of the action is valid when
\be\label{eq;perturbative_condition}
E^+, E^{\times}, {|\vec{k}| \over \bar{F}_{\eta x} } \delta A^+, {|\vec{k}| \over \bar{F}_{\eta x} } \delta A^- , r \ll 1.
\ee

\subsection{Comparison with data}\label{sec:gstar}

A formalism for finding signatures of a generic primordial preferred direction in the CMB has been developed \cite{Ackerman:2007nb,Himmetoglu:2009mk}. In \cite{Ackerman:2007nb} a small direction-dependent contribution to the primordial curvature power spectrum is parametrized by $g_*$ where
\be\label{gstar}
P(\vec{k}) = P_0(k)(1 + g_* \, (\hat{n}\cdot \hat{k})^2)
\ee
and where $\hat{n}$ is some preferred direction in the sky.
It is postulated that $g_*$ will be approximately independent of the scale for modes of astrophysical interest and that parity is still conserved. Parity conservation guarantees the absence of terms with odd powers of $(\hat{n} \cdot \hat{k})$. Contributions proportional to higher powers of $(\hat{n} \cdot \hat{k})^2$ are assumed to be negligible.

Using this formalism, a nonzero value for $g_*$ was found using 5-year WMAP data at the nine sigma level \cite{Groeneboom:2009cb}. The central value found for $g_*$ is $0.29$ for a preferred direction very close to the ecliptic pole.  Since the WMAP scanning strategy is tied to the ecliptic plane, this strongly suggests that the nonzero value of $g_*$ is due to some systematic effect \cite{Hanson:2009gu, Groeneboom:2009cb}. Still, we may reasonably take from the analysis in \cite {Groeneboom:2009cb} an upper bound for $g_*$ of 
\be \label{eq:gstarlimit}
|g_*| < 0.3.
\ee
In \cite{Pullen:2007tu} it is estimated that Planck will be sensitive to values of $|g_*|$ as small as $0.02$. 

Obviously, the gravitational wave power spectrum has not yet been measured, so there is no limit on the analogous parameter, $g_{*\text{grav}}$, for the gravitational wave power spectrum.

\section{Perturbations: odd sector} \label{sec:odd_sector}

As described in section \ref{subsec:decomp}, the quadratic action separates into two uncoupled pieces according to a residual symmetry under parity transformations. We'll therefore analyze the two ``sectors''---which we refer to as ``odd'' and ``even'' for reasons discussed in appendix \ref{app:parameterization}---in different sections.  We start in this section by analyzing the odd sector\footnote{Our odd sector corresponds to the $2d$-vector sector analyzed numerically in \cite{Himmetoglu:2009mk}.} because it is less complicated than the even sector, having only two coupled degrees of freedom (a tensor and a vector degree of freedom) instead of three degrees of freedom as in the even sector. The even sector, which includes the curvature perturbation, contains the most interesting physics; analyzing the odd sector is valuable for extracting $g_{*\text{grav}}$ and as a warm-up for the analysis of the even sector. 

In this section we present the action for the odd sector to quadratic order in gauge-invariant perturbation variables. Then we argue that the form of the action implies that the background is classically stable. Next we diagonalize the kinetic term in the action by defining new perturbation variables in terms of which the kinetic term in the action is canonically normalized. This diagonalization allows us to identify the fields that should be quantized. The Hamiltonian derived from the diagonal form of the action is then separated into a ``free'' part and an ``interacting'' part, and ``in-in'' perturbation theory is used to find the autocorrelations (power spectra) and cross correlations of the vector and tensor degrees of freedom (see \eqref{eq:fieldDefs}) in terms of the preferred direction and the background quantities $H$ and $\Sigma$. The most interesting result in this section is the tensor perturbation power spectrum, given in \eqref{eq:tensor power spectrum}. 

In the odd sector, the action takes the form  
\begin{multline} \label{oddAction}
S^{\text{odd}} = \int d\eta \int {d^3 k \over (2\pi)^3} \Big(  \half \hoddc{'} \hodd{'} + \half \Aoddc{'} \Aodd{'} - \half \hoddc \hodd \left( k^2 - {a'' \over a}  - 4 \rh \alpha'^2 / 3  + \half \dpk \alpha'^2 \left( 2 \rh/3 + 6 \S^2 - {3 \over 2} \dpk \S^2 \right) \right) \\
	- \half \Aoddc \Aodd \left( k^2 - {\dttf \over f} + 2 \S \alpha' {\dtf \over f} + \alpha'^2( 2 \rh - 2 \S + 2 \dpk \rh/3 - \S^2 )\right) 
	\\
	+\left( i \psi_{\vec{k}}' \hoddc \Aodd \left( {{\dtf \over f} + \alpha ' \S + \dpk \alpha ' \S}\right) - i \psi_{\vec{k}}' \hoddc {\Aodd}{'} + \text{h.c.} \right) \Big)
\end{multline}
where
\be
k^2 \equiv \gamma^{i j} k_i k_j =  k_1^2 e^{4 \beta} +  k_2^2 e^{-2 \beta} ,
\ee
\be
\dpk \equiv {k^2{'} \over k^2 \beta'} = { 4 \, k_1^2 e^{4 \beta} - 2 \, k_2^2 e^{-2 \beta} \over  k_1^2 e^{4 \beta} +  k_2^2 e^{-2 \beta} },
\ee
\be \label{eq:priprime}
{\psi_{\vec{k}}' \over \alpha'} \equiv {k_2 e^{-\beta} \over \sqrt{k^2} } \sqrt{\rh}  ,
\ee
and $f'$ denotes the derivative of $f(\bar{\phi}(\eta))$ with respect to conformal time. Without loss of generality we have set $k_3 = 0$ and we have taken the preferred direction (the direction along which the background electric field points) to be $\hat{x}^1$.

By inspection we can see that $\hodd$ and $\Aodd$ decouple when the wave vector is parallel to the preferred direction (so $k_2=0$). This decoupling should be expected due to the enhanced rotational symmetry about the wave vector in this case.

\subsection{Preliminary look at stability}

By design, the kinetic terms are canonically normalized. And in the short wavelength limit ($k \gg a H$), the action simplifies to that of two uncoupled harmonic oscillators; there's no indication of instability in the short wavelength limit. 

Let's consider the case where $k_2=0$ so the wave vector corresponding to a mode points in the preferred direction. In this case, $\psi_{\vec{k}}' = 0$ and $\Delta_{\vec{k}} = 4$. By inspection, one sees that the cross-terms vanish. More explicitly,
\begin{multline}
S^{\text{odd}} \longrightarrow_{k_2 \rightarrow 0} \int d\eta \int {d^3 k \over (2\pi)^3} \Big(  \half \hoddc{'} \hodd{'} + \half \Aoddc{'} \Aodd{'} - \half \hoddc \hodd \left( k^2 - {a'' \over a}  \right) \\
	- \half \Aoddc \Aodd \left( k^2 - {\dttf \over f} + 2 \S \alpha' {\dtf \over f} + \alpha'^2(14 \rh/3 - 2 \S  - \S^2 )\right) \Big).
\end{multline}

When $k_2 \rightarrow 0$ the action for $\hodd$ takes the same form as in the isotropic case. Though the effective mass for $\hodd$ is not real for all time (so naively, there's a tachyon), the important point is that $\hodd / a$, which we assumed to be much less than one in our perturbative expansion of the metric (see \eqref {eq;perturbative_condition}), oscillates with decaying amplitude before horizon crossing, and then remains constant or decays after horizon crossing. In other words, $\hodd \sim a E^{\times} $ never increases faster than $a$, which is consistent with the perturbative expansion. Similarly, given that $2 \S \alpha' {\dtf \over f} + \alpha'^2(14 \rh/3 - 2 \S  - \S^2) \ll {\dttf \over f} $, the long wavelength solution for $\Aodd$ is approximately, $\Aodd \approx C_1 f + C_2 f \int {d \eta \over f^2} $. Now given that $f \approx a^{-2} \approx H^2 \eta^2$, one can see that $ {|\vec{k}| \over \bar{F}_{\eta x} } \delta A^- \sim  (C_1  + {C_2 \over H} a^3) a^{-4} $ (which is decaying) in the long wavelength limit. So clearly the perturbative expansion of the action remains valid when $k_2 = 0$. 

Now let's consider a wave vector that's antiparallel to the preferred direction, so $k_1 = 0$. In this case, $\psi_{\vec{k}}' = \sqrt{\rh}\, \alpha'$ and $\Delta_{\vec{k}} = -2$. Then the effective mass squared for $\hodd$ becomes 
$$
m_{\text{eff}}^2 = k^2 - {a'' \over a} - \alpha'^2(2 \rh + 9 \Sigma^2).
$$
Compared to the isotropic case, the effective mass squared for $\hodd$ receives an additional negative contribution. This suggests that $\hodd$ will grow slightly faster than $a$ outside the horizon. The situation is, of course, complicated by the coupling to $\Aodd$, but all extra terms in the action when $k_1 =0$ compared to the terms present when $k_2=0$ are small. This suggests that any possible growth of the perturbative fields in this case will be very moderate and does not represent an instability. This reasoning will be checked by calculating the power spectra of perturbative fields; we can check that the magnitudes of power spectra do not grow rapidly in time. 

The same situation occurs in the even sector; perturbations clearly do not grow when $k_2 = 0$ and all extra terms in the action when $k_1 =0$ compared to the terms present when $k_2=0$ are small.

\subsection{Diagonalized action}

In general, the canonical quantization of a theory can only proceed once the kinetic interactions have been diagonalized. 
Usually the diagonalization is accomplished by some constant field redefinition.  In our case, we need a time-dependent field redefinition because the ``coefficients'' in the kinetic portions of the action are not constant. (See appendix \ref {ap:diagonal}.)

The kinetic terms can be diagonalized by performing a time-dependent unitary rotation 
\begin{equation}
 \left( \begin{matrix} \hodd \\ \Aodd \end{matrix}  \right) = \left( \begin{matrix} \cos \psi_{\vec{k}}(\eta) & - i \sin \psi_{\vec{k}}(\eta) \\ - i \sin \psi_{\vec{k}}(\eta) & \cos \psi_{\vec{k}}(\eta) \end{matrix} \right)  \left( \begin{matrix} U_1 \\ U_2 \end{matrix}  \right).
\end{equation}
In terms of the rotated fields, $U_i$, the odd-sector action takes the form
\begin{equation} \label{diagonalActionOdd}
S^{\text{odd}} = \int d \eta \int {d^3 k \over (2\pi)^3} \left[ {1 \over 2} \left( \begin{matrix} U_1' \\ U_2' \end{matrix}  \right)^\dagger\left( \begin{matrix} U_1' \\ U_2' \end{matrix}  \right) - {1 \over 2} \left( \begin{matrix} U_1 \\ U_2 \end{matrix}  \right)^\dagger M \left( \begin{matrix} U_1 \\ U_2 \end{matrix}  \right) \right]
\end{equation}
where the Hermitian matrix $M$ is defined
\begin{eqnarray} \label{Mequation}
M &\equiv& \left(k^2 - {1 \over 2}\left({a'' \over a} + {f'' \over f}\right) + \Sigma \alpha'^2 \left({f' \over f \alpha'} -1 - {1 \over2} \Sigma+{3\over2} \Sigma \dpk - {3 \over 8} \Sigma \dpk^2\right) + {1\over 3}\rh \alpha'^2\left(3 + \dpk\right)\right) \mathbb{I} \nonumber \\
&+& \left[\sin(2\psi_{\vec{k}})\sigma_3 - \cos(2\psi_{\vec{k}})\sigma_2\right]\left({\psi_{\vec{k}}' \over \alpha'}\right)\alpha'^2 \left(1 - {f' \over f \alpha'} + \Sigma - {3 \over 2} \Sigma \dpk \right)  \\ 
&+& \left[\cos(2 \psi_{\vec{k}})\sigma_3 + \sin(2\psi_{\vec{k}}) \sigma_2\right] \left({1\over 2} \left({f'' \over f} - {a'' \over a}\right) -\Sigma \alpha'^2\left({f' \over f \alpha'} -1 - {1 \over2} \Sigma-{3\over2} \Sigma \dpk + {3 \over 8} \Sigma \dpk^2\right) -{1\over 3}\rh \alpha'^2\left(5 + {1\over2} \dpk \right)\right) \nonumber
\end{eqnarray}
and where $\mathbb{I} $ is the $2 \times 2$ identity matrix and we have used the following convention for the Pauli matrices
\begin{equation} \label{Pauli}
 \sigma_2 =  \left( \begin{matrix} 0 & - i  \\ i  & 0 \end{matrix} \right) ~~\text{and} ~~  \sigma_3 =  \left( \begin{matrix} 1 & 0  \\ 0  & -1 \end{matrix} \right).
\end{equation}

Physical quantities should not depend on the initial value of $\psi_{\vec{k}}$. Indeed, we will see that correlations of $\hodd$ and $\Aodd$ at a time, $\eta$, calculated using the ``in-in'' formalism of perturbation theory, depend only on the change in $\psi_{\vec{k}}$ after horizon crossing.

\subsection{Correlations using perturbation theory}
In order to calculate correlations, we use the ``in-in'' formalism of perturbation theory, taking the small parameters to be $\epsilon$, $\delta$, $\rh$, and $\Sigma$. As discussed at the end of section \ref{Model} we take
\begin{equation}
\epsilon  = {\alpha'^2 - \alpha''  \over \alpha'^2} \ll 1, \qquad
\delta = { \partial_t^2 H \over 2 H \partial_t H} \ll 1, \qquad
\rh \approx {9 \Sigma / 2} \lesssim \mathcal{O}(\epsilon),\qquad
{\rh' \over \rh \alpha'}\lesssim \mathcal{O}(\epsilon).
\end{equation}
Given these assumptions and the background field equations \eqref{Einstein00} - \eqref{AnisoEinstein} ,
\be
{f ' \over f \alpha'} = -2 + \mathcal{O}(\epsilon), \qquad {f'' \over  f \alpha'^2} = 2 + \mathcal{O}(\epsilon) =  {a'' \over a \, \alpha'^2} 
\qquad \text{and} \qquad \alpha' \approx - {1 \over  \eta} .
\ee

We choose as our free Hamiltonian
\begin{equation} \label{diagonalActionEven}
H^{\text{odd}}_0 \equiv  \int {d^3 k \over (2\pi)^3} \left[ {1 \over 2} \left( \begin{matrix} U_1' \\ U_2' \end{matrix}  \right)^\dagger\left( \begin{matrix} U_1' \\ U_2' \end{matrix}  \right) + {1 \over 2} \left( \begin{matrix} U_1 \\ U_2 \end{matrix}  \right)^\dagger M^{(0)} \left( \begin{matrix} U_1 \\ U_2 \end{matrix}  \right) \right]
\end{equation} 
where
\be
M^{(0)} \equiv  \left( \gamma^{ij}(\eta_0) k_i k_j - {2 \over \eta^2} \right) \mathbb{I}.
\ee

The interaction-picture fields then obey the following equations:
\begin{equation}
{d^2 U_i^I \over d \eta^2} + \left(\gamma^{ij}(\eta_0) k_i k_j - {2 \over \eta^2} \right) U_i^I = 0.
\end{equation}
Each of these fields can be expanded in terms of time-independent creation and annihilation operators as,
\begin{equation}
U_i^I(\vec{x},\eta)= \int {d^3 k \over (2\pi)^{3}} e^{ i k_j x^j}U_i^I(\vec{k},\eta)= \int {d^3 k \over (2\pi)^{3}} \left(e^{i k_i x^i}\chi^{(0)}(k_{\eta_0},\eta)\hat{a}^i_{\vec{k}}+ e^{-i k_i x^i}\chi^{(0)}{}^{*}(k_{\eta_0},\eta)(\hat{a}^i_{\vec{k}})^\dagger\right),
\end{equation}
where the canonically normalized mode functions are
\begin{equation}
\chi^{(0)}(k,\eta) ={e^{-i k \eta} \over \sqrt{2 k}}\left(1- {i \over k \eta}\right)
\end{equation}
and where the commutation relations of the creation and annihilation operators are
\begin{eqnarray}
\left[\hat{a}^i_{\vec{k}},(\hat{a}^j_{\vec{q}})^\dagger\right] = (2 \pi)^3 \delta^{ij} \delta (\vec{k} - \vec{q}) ~~ \text{and}~~\left[\hat{a}^i_{\vec{k}},\hat{a}^j_{\vec{q}}\right] =  0.
\end{eqnarray}
Here, 
\be
k_{\eta_0} \equiv \sqrt{\gamma^{i j}(\eta_0) k_i k_j }.
\ee

If we choose $\beta_0 = 0$ then  $\gamma^{i j}(\eta_0) = \delta^{i j}$. But then if $\beta$ changes during inflation, the coordinates at the end of inflation will not be isotropic. On the other hand, if we choose $\beta_0$ so that $\beta = 0$ at the end of inflation (when the dynamics returns to being isotropic), then the coordinates at the end of inflation will be isotropic.  The latter choice is more convenient.

Using the results of the previous section and the form of the matrix $M$ in \eqref{Mequation}, the interaction-picture Hamiltonian takes the form
\begin{equation}
H_I(\eta) = \int {d^3k \over (2\pi)^3}\left(  {1 \over 2} \left( \begin{matrix} U_1^I \\ U_2^I \end{matrix}  \right)^\dagger M^{(1)} \left( \begin{matrix} U_1^I \\ U_2^I \end{matrix} \right) \right)
\end{equation}
where,
\be
M^{(1)} = M - M^{(0)} = f_1(\eta, \vec{k}) \mathbb{I} + \left[\sin(2\psi_{\vec{k}})\sigma_3 - \cos(2\psi_{\vec{k}})\sigma_2\right] f_2(\eta, \hat{k}) + \left[\cos(2 \psi_{\vec{k}})\sigma_3 + \sin(2\psi_{\vec{k}}) \sigma_2\right] f_3(\eta, \hat{k})
\ee
and we have defined
\begin{align}
f_1(\eta, \vec{k}) &\equiv  (\gamma^{i j}(\eta)-\gamma^{ij}(\eta_0))k_i k_j - {1 \over 2}\left({a'' \over a} + {f'' \over f} - {4 \over \eta^2 }\right) \nonumber \\ &+ \Sigma \alpha'^2 \left({f' \over f \alpha'} -1 - {1 \over2} \Sigma+{3\over2} \Sigma \dpk - {3 \over 8} \Sigma \dpk^2\right) 
+ {1\over 3}\rh \alpha'^2\left(3 + \dpk\right) \\ \label{eq:f1} 
f_2(\eta, \hat{k}) &\equiv \left({\psi_{\vec{k}}' \over \alpha'}\right)\alpha'^2 \left(1 - {f' \over f \alpha'} + \Sigma - {3 \over 2} \Sigma \dpk \right) \\  
f_3(\eta, \hat{k}) &\equiv \left({1\over 2} \left({f'' \over f} - {a'' \over a}\right) -\Sigma \alpha'^2\left({f' \over f \alpha'} -1 - {1 \over2} \Sigma-{3\over2} \Sigma \dpk + {3 \over 8} \Sigma \dpk^2\right) -{1\over 3}\rh \alpha'^2\left(5 + {1\over2} \dpk \right)\right). \label{eq:f3} 
\end{align}

Our convention for the correlations of the fields will be
\begin{equation}
\left<U_i(\vec{k},\eta) U_j(\vec{q},\eta) \right> = C_{ij}(\vec{k}, \eta) (2\pi)^3 \delta(\vec{k}+\vec{q}),
\end{equation}
where the power spectra are the diagonal entries of the matrix $C_{ij}$.  Using \eqref{perturbedpower}, the correlations can be written as 
\begin{equation}
\langle U_i(\pv, \eta ) \, U_j(\qv, \eta ) \rangle = \langle U^I_i(\pv, \eta) U^I_j(\qv, \eta) \rangle + i \int^\eta d \eta' \langle [H_I(\eta'), U^I_i(\pv, \eta) U^I_j(\qv, \eta) ]\rangle + \ldots
\end{equation}
More explicitly, the correlations take the form
\begin{equation} \label{correlations}
C_{i j}(\pv, \eta) = |\chi^{(0)}(p_{\eta_0}, \eta)|^2 \delta_{i j} + i \int^\eta d \eta'  M^{(1)}_{i j} (\pv, \eta')I_{p_{\eta_0}}(\eta',\eta) + \ldots
\end{equation}
where
\begin{equation}
I_{p}(\eta',\eta) =  \left(  (\chi^{(0)}(p, \eta') \chi^{(0) {*}}(p, \eta))^2 -   (\chi^{(0) {*}}(p, \eta') \chi^{(0)}(p, \eta))^2 \right).
\end{equation} 
It is clear from this formula that the zeroth-order power spectra of the fields $U_i$ are isotropic and scale invariant and that the cross-correlation vanishes.   Here it's convenient to define the function
\be
\tilde{I}(p \eta',p \eta) \equiv i p^2 I_{p}(\eta',\eta) 
\ee
 where,
\begin{equation}\label{eq:Itilde}
\tilde{I}(x,y) = \left(\frac{1}{2 x^2 y^2}-\frac{1}{2 x^2}+\frac{2}{x y}-\frac{1}{2 y^2}+\frac{1}{2}\right) \sin (2 x-2 y)+\left(\frac{1}{x^2 y}-\frac{1}{x y^2}+\frac{1}{x}-\frac{1}{y}\right) \cos(2 x-2 y).
\end{equation}

Solving for the correlations of the variables $\hodd$ and $\Aodd$ in terms of the correlations of the rotated variables $U_i$, we find
\begin{eqnarray}
P_{\hodd}(\vec{p}) &=& \cos^2\psi_{\vec{p}} \,C_{11}(\vec{p}) + \sin^2\psi_{\vec{p}} \,C_{22}(\vec{p}) + {i\over 2} \sin(2\psi_{\vec{p}})(C_{12}(\vec{p}) - C_{21}(\vec{p})) \label{eq:rotated1}\\
P_{\Aodd}(\vec{p}) &=& \sin^2\psi_{\vec{p}} \,C_{11}(\vec{p}) + \cos^2\psi_{\vec{p}} \,C_{22}(\vec{p}) - {i\over 2} \sin(2\psi_{\vec{p}})(C_{12}(\vec{p}) - C_{21}(\vec{p})) \\
-C_{\Aodd \hodd}(\vec{p}) = C_{\hodd \Aodd}(\vec{p}) &=& \cos^2\psi_{\vec{p}}\, C_{12}(\vec{p}) + \sin^2\psi_{\vec{p}} \,C_{21}(\vec{p}) + {i\over 2} \sin(2\psi_{\vec{p}})(C_{11}(\vec{p}) - C_{22}(\vec{p})) \label{eq:rotated3}
\end{eqnarray}
where we have used the fact that $\psi_{-\vec{k}} = - \psi_{\vec{k}}$.  All of the above correlations, and $\psi_{\vec{p}}$, are functions of time.  It is understood that these expressions are evaluated at the end of inflation.

From here on, we will use the short hand notation
\be
p = p_{\eta_0}.
\ee

Using \eqref {correlations} and the expression for $M^{(1)}$, the power spectra and correlations are given more explicitly by, 
\begin{align} \label{eq:Ph}
P_{\hodd}(\vec{p},\eta) &=  |\chi^{(0)}(p, \eta)|^2+ p^{-2} \bigg\lbrace \int^\eta f_1(\eta', \vec{p})\, \tilde{I}(p \eta',p \eta) d\eta' + \left(\int^\eta \sin\left( 2\psi_{\vec{p}}(\eta') - 2\psi_{\vec{p}}(\eta)\right) \, f_2(\eta', \hat{p}) \, \tilde{I}(p \eta',p \eta) d\eta'\right) \nonumber \\ & \qquad + \left(\int^\eta \cos\left( 2\psi_{\vec{p}}(\eta') - 2\psi_{\vec{p}}(\eta)\right) \, f_3(\eta', \hat{p}) \, \tilde{I}(p \eta',p \eta)  d\eta' \right) \bigg\rbrace + \ldots \\
P_{\Aodd}(\vec{p},\eta) &=  |\chi^{(0)}(p, \eta)|^2+ p^{-2} \bigg\lbrace \int^\eta f_1(\eta', \vec{p})\, \tilde{I}(p \eta',p \eta) d\eta' - \left(\int^\eta \sin\left( 2\psi_{\vec{p}}(\eta') - 2\psi_{\vec{p}}(\eta)\right) \, f_2(\eta', \hat{p}) \, \tilde{I}(p \eta',p \eta) d\eta'\right) \nonumber \\ & \qquad - \left(\int^\eta \cos\left( 2\psi_{\vec{p}}(\eta') - 2\psi_{\vec{p}}(\eta)\right) \, f_3(\eta', \hat{p}) \, \tilde{I}(p \eta',p \eta)  d\eta' \right) \bigg\rbrace + \ldots \\
C_{\hodd \Aodd}(\vec{p},\eta) &= i p^{-2} \bigg\lbrace  \int^\eta \cos\left( 2\psi_{\vec{p}}(\eta') - 2\psi_{\vec{p}}(\eta)\right) \, f_2(\eta', \hat{p}) \, \tilde{I}(p \eta',p \eta)  d\eta'  \\  \label{eq:ChA}
	& \qquad - \int^\eta \sin\left( 2\psi_{\vec{p}}(\eta') - 2\psi_{\vec{p}}(\eta)\right) \, f_3(\eta', \hat{p}) \, \tilde{I}(p \eta',p \eta)  d\eta'  \bigg\rbrace  + \ldots = - C_{\Aodd \hodd}(\vec{p},\eta). 
\end{align}
It's clear from the expression above that the correlations are functions only of the change in the angle $\psi_{\vec{p}}$.

\subsection{Discussion}

We are interested primarily in \emph{direction-dependent} modifications to the power spectra---\emph{i.e.}, modifications of the power spectra that depend on the direction of the wave vector, not just its magnitude. Non-direction-dependent effects will modify spectral indices, but such effects cannot be disentangled experimentally as due to primordial anisotropy. In principle, one could use our method to calculate spectral indices and, for example, relate them to the size of the direction-dependent effects.

The largest direction-dependent contribution comes from the piece involving $f_2$. The contribution is given by,
\begin{equation} \label{eq:odd_gstar_contribution}
p^{-2} \left(\int^\eta \sin\left( 2\psi_{\vec{p}}(\eta') - 2\psi_{\vec{p}}(\eta)\right) \, f_2(\eta', \hat{p}) \, \tilde{I}(p \eta',p \eta) d\eta'\right) \approx - {(a H)^2 \over p^3} \left( \cos[2 {\psi'_{\vec{p}} \over \alpha'} \log(a H / p)] - 1 \right)
\end{equation}
assuming ${\psi'_{\vec{p}} \over \alpha'} $ is approximately constant throughout inflation, where we've used the fact that $\left( 1 - {f' \over f \alpha' }\right) \approx 3$ and the relevant integral is calculated in appendix  \ref{app:integrals}.  Modes of astrophysical interest crossed the horizon about sixty $e$-folds before the end of inflation, so for such modes, $\log(a H / p) \approx 60 $. 

When 
\be
f(\phi) = \exp \left\lbrace 2 c \kappa \int  \left( {\partial_\phi V \over \kappa V} \right)^{-1} d \phi \right\rbrace,
\ee
for $c-1 \sim \mathcal{O}(1)$ we found that $\rh \approx {3 (c-1) \over 2 c} \epsilon $ during the anisotropic period of expansion. If the anisotropic period of expansion were to last all sixty $e$-folds before the end of inflation, then we should expect order one direction-dependent corrections to the gravitational wave power spectrum for inflationary scenarios in which $\sqrt{\epsilon} \gtrsim {1 \over 60}$. Such values of $\epsilon$ can easily be realized in large-field inflationary models. This analytic result seems to confirm the numerical findings in \cite{Himmetoglu:2009mk}.  

Demanding that the direction-dependent effect on the gravitational wave power spectrum for modes of astrophysical interest is less than, say, about $30 \%$ would mean that the argument of the cosine function in \eqref {eq:odd_gstar_contribution} is small so that the cosine can be expanded in a Taylor series. In this case the power spectrum for $\hodd$ is approximately,
\be\label{eq:tensor power spectrum}
P_{\hodd}(\vec{p},\eta) \approx {(a H)^2 \over 2p^3}( 1 +  \left(2 {\psi'_{\vec{p}} \over \alpha'} \log(a H / p) \right)^2) \approx {(a H)^2 \over 2p^3}\left( 1 + 4 \rh (\log(a H / p))^2(1 - (\hat{n}\cdot\hat{p})^2)\right) .
\ee
where $\hat{n}$ is the preferred direction. Thus we may identify
\be\label{eq:g_star_grav}
g_{*\text{grav}} \approx -4 \rh (\log(a H / p))^2 \approx -18 \Sigma (\log(a H / p))^2.
\ee
Note that $g_{*\text{grav}}$ is nearly (though not exactly) scale invariant for modes of astrophysical interest.

Imposing a limit like $|g_{* \text{grav}}| < 0.3$ for modes of astrophysical interest corresponds to a limit on $\rh$ like 
\be 
 \rh|_{\text{average after horizon crossing}} \lesssim 10^{-4} \qquad \text{when} \qquad |g_{* \text{grav}}| < 0.3.
\ee


\section{Perturbations: even sector}\label{sec:even_sector}
The even-sector action is much more complicated than that of the odd sector. This sector contains three dynamical degrees of freedom that, in the isotropic limit, transform as a scalar, vector and tensor under rotations.  This sector is further complicated by additional nondynamical scalar variables.  

As in the previous section, we begin in this section by diagonalizing the kinetic part of the quadratic action. This process is more complicated for the three dynamical degrees of freedom in this (even) sector than for the two of the odd sector, and the smallness of certain background quantities must be exploited; we eventually work in the limit $\rh \ll \e \ll 1$, which is confirmed to be a sensible limit at the end of the calculation. As in the odd-sector calculation,  we quantize and use ``in-in'' perturbation theory to calculate power spectra and cross correlations of the scalar, vector, and tensor degrees of freedom. The most interesting results in this section are the scalar perturbation power spectrum \eqref{eq:scalar power spectrum} and corresponding value for $g_*$ \eqref{gstarR}, and also the ratio of the direction-dependent correction to the scalar power spectrum over that of the tensor power spectrum \eqref{eq:consistency condition}.

Instead of presenting the entire quadratic action (as we did in \eqref{oddAction} for the odd sector), here we present the action to lowest order in $\delta$, $\epsilon$, $\rh$, and $\Sigma$. We expand the action assuming that $\rh$,  $\Sigma$, and  $\rh' / \rh$  are order $\epsilon$ or smaller. For simplicity, we first present the action to lowest order before elimination of the auxiliary fields $\Phi$ and $\Psi$. (See appendix \ref {app:parameterization}  for the definitions of $\Phi$ and $\Psi$.) The action can be written
\be
S{^\text{even}} = \int d \eta \int {d^3 k \over (2 \pi)^3} \left[ \H^{\dagger} \Mone \H + \P^{\dagger} \Q \H + \H^{\dagger} \Q^{\dagger} \P + \P^{\dagger} \Mtwo \P \right]
\ee
where the vectors $\H$ and $\P$ are defined by
\be
\H = \left( \begin{matrix} \hev {'} \\ \Aev {'} \\ \rev {'} \\ \hev \\ \Aev \\ \rev \end{matrix} \right) \qquad \P = \left( \begin{matrix} \Phi \\ \Psi  \end{matrix} \right)
\ee
and the matrices $\Mone$, $\Mtwo$, and $\Q$ are given by
\be \label{eq:M1A}
\Mone = \left( \begin{matrix}  
\half  & 0 & 0 & 0 & 0 & 0 \\ 
0 & \half & 0  & - i \ktwo  & 0 & - i 2 \sqrt{2} {a  \over \kappa z} \ktwo \\ 
0 & 0 & \half  & 0 & 0 & 0 \\ 
0 & i \ktwo & 0  & \alpha'^2 -{k^2 \over 2} & 2 i \ktwo \alpha' & 2 \sqrt{2} {a  \over \kappa z} \ktwo^2 \\ 
0 & 0 & 0 & -2 i \ktwo \alpha'  & \alpha'^2-{k^2 \over 2}  & - i 4 \sqrt{2} {a \alpha' \over \kappa z} \ktwo \\ 
0 &  i 2 \sqrt{2} {a \over \kappa z} \ktwo  & 0  & 2 \sqrt{2} {a \over \kappa z}\ktwo^2  &  i 4 \sqrt{2} {a \alpha' \over \kappa z} \ktwo  & \half {z'' \over z} -{k^2 \over 2} + 16 {a^2 \over \kappa^2 z^2} \ktwo^2  - 8 {a^2 \alpha'^2 \over \kappa^2 z^2} \rh   \end{matrix} \right) 
+ \mathcal{O}(\epsilon)
\ee
\\
\\
\begin{multline}
\Mtwo =  \left( \begin{matrix}  {a^2 \over \kappa^2 }\ktwo^2   & - {3 a^2 \over \kappa^2 }(\ktwo^2+ {(2/3) \rh \alpha'^2 })  -  {3 \over 2} z^2 \alpha'^2 \\ 
- {3 a^2 \over \kappa^2 }(\ktwo^2+ {(2/3) \rh \alpha'^2 })  -  {3 \over 2} z^2 \alpha'^2  & 
 {9 a^2 \over \kappa^2 }( \ktwo^2 +  { (2/3) \rh \alpha'^2 })- {\kappa^2 z^2 \over a^2} )+  {3 z^2 \alpha'^2 \over 2}(1  +   {2 z' \over \alpha' z})  \end{matrix} \right) \\
+ {a^2  k^2  \over \kappa^2 } \left( \begin{matrix}  0  & -  (1 - { \dpk \Sigma \over  4 })  \\ 
-  (1 - { \dpk \Sigma \over  4 })   & 
    ( 1 +{ \dpk \Sigma \over 2 } - {\kappa^2 z^2 \over 2 a^2} )  \end{matrix} \right) + \mathcal{O}(\epsilon^2)
\end{multline}
\\
\be
\Q = \left( \begin{matrix}  
0  & i {a \over \sqrt{2} \kappa } \ktwo  & 0 & \sqrt{2}{a \over \kappa} \ktwo^2 & -i {a \over \sqrt{2} \kappa } \ktwo \alpha' &4 {a^2  \over \kappa^2 z}\ktwo^2  \\ 
0  & 0 & 0 & - 3 \sqrt{2} {a  \over \kappa} \ktwo^2 + {a k^2 \S \over 4 \sqrt{2} \kappa} (\dpk - 4) & 0 & \half k^2 z -12 {a^2  \over \kappa^2 z} \ktwo^2 
 \end{matrix} \right) 
 +\mathcal{O}(\epsilon^{3/2})
\ee
and $\ktwo$ is as in \eqref {eq:priprime}.
Note here the identity
\be
\alpha'^2(\dpk - 4 ) \rh = -4 \ktwo^2.
\ee

Solving the (constraint) equations of motion derived by varying the action with respect to $\Phi$ and $\Psi$ and plugging the constraint equations back into the action leads to the action in terms of the three dynamical fields:
\be \label{eq:evenAction}
S{^\text{even}} = \int d \eta \int {d^3 k \over (2 \pi)^3} \left[ \H^{\dagger} \left( \Mone -\Q^{\dagger} \Mtwo^{-1} \Q \right) \H  \right].
\ee

Keep in mind that $\ktwo$ is a direction-dependent quantity that varies from zero to plus or minus $\sqrt{\rh }$, depending on the orientation of the wave vector with respect to the preferred direction. The bottom right element of $\Mone$, representing (minus) the effective mass for $\rev$, is $\half ({z'' \over z} - k^2 )$ in the isotropic limit.  So if, for example, $\rh$ is order ${\kappa^2 z^2 \over a^2} = \mathcal{O}(\epsilon)$ then we should expect a very dramatic direction-dependent effect on the curvature perturbation power spectrum, because the direction-dependent term would be on the same order as the normal, isotropic term (at least in the long wavelength limit). In fact, assuming that taking into account the $\Q^{\dagger} \Mtwo^{-1} \Q$ correction to $\Mone$ and properly diagonalizing the kinetic term in the action would not weaken the direction-dependent effect on the power spectrum, we can get a rough limit on the average value of  $\rh / (\kappa^2 z^2 / a^2)$ during inflation, after horizon crossing.  Based on the argument of section \ref {sec:gstar}, we may take a $30 \%$ direction-dependent contribution to curvature perturbation power spectrum to be an upper limit. Noting that ${z'' \over z} =  \alpha'^2(2  + \mathcal{O}(\epsilon, \delta))$, the $30 \%$ limit translates roughly to\footnote{The first equality can be seen from equations \eqref {eq:phiprime} and \eqref{eq:z}, given that $\rh$ must be small compared to $\kappa^2 z^2 / a^2$. } 
\be \label{eq:approx_constraint}
{ {\rh a^2} \over  \kappa^2 z^2 }\Big|_{\text{average}} \approx { {\rh } \over 2 \epsilon }\Big|_{\text{average}}  < 10^{-2} \qquad (\text{approximate}).
\ee

Given phenomenological constraints, it is therefore most interesting to consider scenarios in which $\rh \ll \epsilon$. Taking 
\be
\rh \sim (9 / 2) \Sigma \ll \epsilon,
\ee
by inspection one can see that in the long wavelength limit, 
\be
\Q^{\dagger} \Mtwo^{-1} \Q = \mathcal{O}( \rh / \epsilon )
\ee
and
\be \label{MoneSimp}
\Mone = 
\left( \begin{matrix}  
\half  & 0 & 0 & 0 & 0 & 0 \\ 
0 & \half & 0  & 0  & 0 & - i 2 \sqrt{2} {a  \over \kappa z} \ktwo \\ 
0 & 0 & \half  & 0 & 0 & 0 \\ 
0 & 0& 0  & \alpha'^2 -{k^2 \over 2} & 0 & 0 \\ 
0 & 0 & 0 & 0 &  \alpha'^2 -{k^2 \over 2} &- i 4 \sqrt{2} {a \alpha' \over \kappa z} \ktwo \\ 
0 &  i 2 \sqrt{2} {a \over \kappa z} \ktwo  & 0  & 0  &  i 4 \sqrt{2} {a \alpha' \over \kappa z} \ktwo  & \half {z'' \over z} -{k^2 \over 2}   \end{matrix} \right) 
+ \mathcal{O}(\epsilon, \rh /\epsilon).
\ee
We will find, with a careful analysis in the $\rh \ll \epsilon$ limit, that the actual constraint on $\rh$ is much stronger than the approximate constraint in \eqref {eq:approx_constraint}.  Thus the $\rh \ll \epsilon$ approximation is valid.

\subsection{Diagonalizing the action}
Once again, the resulting kinetic terms are not diagonalized and canonical quantization cannot proceed. In the  $\rh \ll \epsilon \ll 1$  limit, the kinetic terms can be diagonalized by performing a time-dependent unitary rotation 
\begin{equation}
 \left( \begin{matrix} \rev \\ \Aev \end{matrix}  \right) = \left( \begin{matrix} \cos \theta_{\vec{k}}(\eta) & - i \sin \theta_{\vec{k}}(\eta) \\ - i \sin \theta_{\vec{k}}(\eta) & \cos \theta_{\vec{k}}(\eta) \end{matrix} \right)  \left( \begin{matrix} U_1 \\ U_2 \end{matrix}  \right),
\end{equation}
where
\begin{equation}
\theta_{\vec{k}}'(\eta) \equiv - 2 \sqrt{2} {a  \over \kappa z} \ktwo =   - 2 \sqrt{2} {a  \over \kappa z}\left( {k_2 e^{-\beta} \over \sqrt{k^2} } \sqrt{\rh} \alpha' \right)
\end{equation}
and where $\ktwo$ is the rotation angle in the odd sector, given by \eqref {eq:priprime}.  The rotation of  $\rev$ and $\Aev$ occurs on a much faster timescale than that of $\hodd$ and $\Aodd$ since $\ktwo = \mathcal{O}(\sqrt{\rh})$ and $\theta_{\vec{k}}' = \mathcal{O}(\sqrt{\rh / \epsilon})$.

In terms of these rotated fields the even action takes the form
\begin{equation} \label{diagonalActionEven2}
S^{\text{even}} = \int d \eta \int {d^3 k \over (2\pi)^3} \left[ \half \hev{'} {\hev}{^*}{'} -\half \left(k^2 - 2\alpha'^2\right) \hev {\hev}{^*} 
+  {1 \over 2} \left( \begin{matrix} U_1' \\ U_2' \end{matrix}  \right)^\dagger\left( \begin{matrix} U_1' \\ U_2' \end{matrix}  \right) - {1 \over 2} \left( \begin{matrix} U_1 \\ U_2 \end{matrix}  \right)^\dagger M \left( \begin{matrix} U_1 \\ U_2 \end{matrix}  \right) + \ldots \right]
\end{equation}

where the Hermitian matrix $M$ is defined
\begin{equation}
M \equiv \left(k^2 - 2\alpha'^2\right) \mathbb{I} 
+ \left[\sin(2 \theta_{\vec{k}})\sigma_3 - \cos(2\theta_{\vec{k}}) \sigma_2\right]\left(3{\theta_{\vec{k}}' \over \alpha'}\right) \alpha'^2
\end{equation}
up to corrections of order $\epsilon$, $\delta$, and $\rh / \epsilon$.\footnote{ Recall that,  \emph{e.g.}, $z''/2z = \alpha'^2 + \mathcal{O}(\epsilon, \delta)$ and $z'/z = \alpha' + \order(\epsilon, \delta)$.}  We've used the same convention for Pauli matrices as in Eq. \eqref{Pauli} and, again, $\mathbb{I} $ is the $2 \times 2$ identity matrix.

\subsection{Correlations using perturbation theory}
The analysis of correlations of dynamical fields in this sector will be very similar to that of the odd sector, up to minus signs and replacing $\psi_{\vec{k}}$ with $\theta_{\vec{k}}$.  It should be noted that the largest direction-dependent corrections to correlations in the odd sector are order $\sqrt{\rh}$, whereas here we're working to order $\sqrt{\rh / \epsilon}$ assuming $\rh \ll \epsilon$. It therefore should be unsurprising that the autocorrelation of the gravitational wave amplitude, $\hev$, has no anisotropic contribution at $\mathcal{O}(\sqrt{\rh / \epsilon})$.   The same can be said of the cross-correlation between $\hev$ and $\Aev$.  

Considering now only terms up to order $\sqrt{\rh / \epsilon}$ given $\rh \ll \epsilon$, we choose as our free Hamiltonian,
\begin{equation} \label{diagonalActionEven3}
H^{\text{even}}_0 \equiv  \int {d^3 k \over (2\pi)^3} \left[\half \hev{'} {\hev}{^*}{'} + {1 \over 2} \left( \begin{matrix} U_1' \\ U_2' \end{matrix}  \right)^\dagger\left( \begin{matrix} U_1' \\ U_2' \end{matrix}  \right)  +\half  \left( \gamma^{ij}(\eta_0) k_i k_j - {2 \over \eta^2} \right) \hev {\hev}{^*}  + {1 \over 2} \left( \begin{matrix} U_1 \\ U_2 \end{matrix}  \right)^\dagger M^{(0)} \left( \begin{matrix} U_1 \\ U_2 \end{matrix}  \right) \right]
\end{equation} 
where
\be
M^{(0)} \equiv  \left( \gamma^{ij}(\eta_0) k_i k_j - {2 \over \eta^2} \right) \mathbb{I}.
\ee
The interaction-picture fields then obey the following equations,
\begin{equation}
{d^2 U_i^I \over d \eta^2} + \left(\gamma^{ij}(\eta_0) k_i k_j - {2 \over \eta^2} \right) U_i^I = 0.
\end{equation}
As in section \ref {sec:odd_sector}, the fields can be expanded into appropriately normalized mode functions and time-independent creation and annihilation operators.  Dropping terms of order $\epsilon$, $\rh / \epsilon$, $\delta$ or higher (including terms with coefficients $(\gamma^{i j}(\eta)-\gamma^{ij}(\eta_0))k_i k_j $) the interaction-picture Hamiltonian takes the form
\begin{equation}
H_I(\eta) = \int {d^3k \over (2\pi)^3}\left(  {1 \over 2} \left( \begin{matrix} U_1^I \\ U_2^I \end{matrix}  \right)^\dagger M^{(1)} \left( \begin{matrix} U_1^I \\ U_2^I \end{matrix} \right) \right)
\end{equation}
where
\begin{equation}
M^{(1)} = M - M^{(0)} = 
3  \left[\sin(2 \theta_{\vec{k}})\sigma_3 - \cos(2\theta_{\vec{k}}) \sigma_2\right] \left({\theta_{\vec{k}}' \over \alpha'}\right) \alpha'^2.
\end{equation}
After computing correlations of the rotated variables using the ``in-in'' formalism, we can find the correlations of the unrotated variables using the equations analogous to equations \eqref {eq:rotated1} - \eqref {eq:rotated3}.

The correlations are approximately given by
\begin{align} \label{eq:Pr1}
P_{\rev}(\vec{p},\eta) &\approx  |\chi^{(0)}(p, \eta)|^2 + p^{-2} \left(\int^\eta \sin\left( 2\theta_{\vec{p}}(\eta') - 2\theta_{\vec{p}}(\eta)\right) \, 3 { \theta'_{\vec{p}}(\eta') \over \alpha'(\eta') } \alpha{'}^2(\eta')\, \tilde{I}(p \eta',p \eta) d\eta'\right) \\
P_{\Aev}(\vec{p},\eta) &\approx  |\chi^{(0)}(p, \eta)|^2 - p^{-2} \left(\int^\eta \sin\left( 2\theta_{\vec{p}}(\eta') - 2\theta_{\vec{p}}(\eta)\right) \,3 { \theta'_{\vec{p}}(\eta') \over \alpha'(\eta') } \alpha{'}^2(\eta')\, \tilde{I}(p \eta',p \eta) d\eta'\right) \\
C_{\rev \Aev}(\vec{p},\eta) = - C_{\Aev \rev}(\vec{p},\eta) &\approx i p^{-2} \bigg\lbrace  \int^\eta \cos\left( 2\theta_{\vec{p}}(\eta') - 2\theta_{\vec{p}}(\eta)\right) \, 3 { \theta'_{\vec{p}}(\eta') \over \alpha'(\eta') } \alpha{'}^2(\eta') \, \tilde{I}(p \eta',p \eta)  d\eta'   \bigg\rbrace   .
\end{align}
where $\tilde{I}$ is defined in \eqref {eq:Itilde}.

Assuming $\rh$ and ${\kappa \phi' \over \alpha'} = {z \over \kappa a}$ are nearly constant during inflation, as in the scenarios we described in section \ref {Model}, then 
\be
\theta_{\vec{p}}(\eta) \approx { \theta'_{\vec{p}} \over \alpha' } \, \alpha(\eta) 
\ee
and we may estimate the relevant integral as in appendix \ref {app:integrals}. Then we see that 
\begin{align} \label{eq:Pr}
P_{\rev}(\vec{p},\eta) &\approx  {(a H)^2 \over 2p^3}\left(  1 - 2 \left( \cos \left(\left(2 { \theta'_{\vec{p}} \over \alpha' }\right) \log(a H / p) \right) - 1 \right)
 \right)  \\ 
P_{\Aev}(\vec{p},\eta) &\approx  {(a H)^2 \over 2p^3}\left(  1 + 2 \left( \cos \left(\left(2 { \theta'_{\vec{p}} \over \alpha' }\right) \log(a H / p) \right) - 1 \right)
 \right)  \\
C_{\rev \Aev}(\vec{p},\eta) = - C_{\Aev \rev}(\vec{p},\eta) &\approx  i {(a H)^2 \over p^3} \sin \left(\left(2 { \theta'_{\vec{p}} \over \alpha' }\right) \log(a H / p)\right), 
\end{align}
where ${ \theta'_{\vec{p}} \over \alpha' } $ should be taken as the average value after horizon crossing.

Now $g_*$, the parameter that characterizes the effect of a preferred direction on the CMB power spectrum, is roughly given by 
\be \label{eq:g_star_scalar}
|g_*| \approx - 2 \left( \cos \left(\left(2 { \theta'_{\vec{p}} \over \alpha' }\right) \log(a H / p)\right) - 1 \right)\Big|_{\text{max}}.
\ee

The maximal value of ${ \theta'_{\vec{p}} \over \alpha' } $ for a given wave vector is approximately $ 2 \sqrt{ {\rh \over \epsilon}}$. So even if $\rh / \epsilon$ is, say, order $10^{-4}$, the argument of the cosine in \eqref {eq:g_star_scalar} could be significant for modes of astrophysical interest because for such modes $\log(a H / p) \approx 60 $. It's then clear that $|g_*|$ could be order one even for very small values of $\Sigma$ and $\rh$. 

Let's suppose that $\rh$ is small enough to satisfy the $|g_*| < 0.3$ bound of section \ref {sec:gstar}. Then the cosine in \eqref{eq:Pr} can be expanded in a Taylor series to give
\be\label{eq:scalar power spectrum}
P_{\rev}(\vec{p},\eta) \approx  {(a H)^2 \over 2p^3}\left(  1 + 16 {\rh \over \epsilon} {(\log(a H / p))}^2 (1 - (\hat{n}\cdot \hat{p})^2) \right),
\ee
where $\hat{n}$ is the preferred direction, and therefore 
\be \label{gstarR}
g_* \approx  -16 {\rh \over \epsilon} \left(\log\left({a H \over p}\right)\right)^2 \approx -72 {\Sigma \over \epsilon} \left(\log\left({a H \over p}\right)\right)^2.
\ee

Note that $g_*$ is negative, as is $g_{* \text{grav}}$ (see equation \eqref {eq:g_star_grav}). A negative $g_*$ means that, for a given scale, power is minimized in the preferred direction. We can understand this general feature in the following way: the pressure contributed by the background electric field slows the expansion of the direction along which the electric field points. In other words, expansion is slower along the preferred direction. Generically the power in primordial perturbations increases in proportion to the Hubble parameter squared; the faster the expansion, the more quickly quantum fluctuations are stretched into ``classical'' perturbations. Since the power of primordial perturbations increases with the Hubble parameter, squared, and since in our scenario the space-time is expanding most slowly in the preferred direction, we might expect that the power of perturbations with wave vectors parallel to the preferred direction will be smaller than the power of perturbations with wave vectors in any other direction. We predict that, generically, models in which a preferred direction expands more rapidly/slowly than other directions will lead to positive/negative values of $g_*$.

The limit $|g_*| < 0.3$ translates into a limit on the average value of ${\rh \over \epsilon} $ during inflation (after horizon-crossing) for modes of astrophysical interest:
\be
{\rh \over \epsilon}\Big|_{\text{average after horizon crossing}} < {3 \over 160 \, (60)^2}.
\ee
Since $\rh$ is assumed to be essentially constant during inflation (as is $\hat{\rho}_\phi$), the limit can be written,
\be
{\hat{\rho}_A \over \hat{\rho}_\phi  \epsilon} \Big|_{\text{average after horizon crossing}} \lesssim 10^{-6}.
\ee
The measurement of $g_*$ puts a very stringent constraint on the ratio of vector field energy density to the inflaton energy density.   At the same time, we see that even a very small $U(1)$ gauge field energy density during inflation could lead to a significant direction-dependent effect on the curvature perturbation power spectrum.

Supposing that $\rh \ll \epsilon$, as we've just seen must be the case in order to comply with observation, the ratio of the gravitational wave power spectrum ($P_T$) to the scalar power spectrum ($P_S$) is approximately\footnote{In the last equality we used equations \eqref {eq:phiprime} and \eqref{eq:z}, given that $\rh$ must be small compared to $\kappa^2 z^2 / a^2$.
}$^,$\footnote{What are identified as tensor perturbations are the amplitudes of the transverse, traceless (TT) part of $\delta g_{i j} / a^2$. We defined $\delta g_{i j, TT} / a^2 = 2 E_{i j}$, thus the extra factor of $2^2$.}
\be 
{P_T \over P_S} = 4 { P_{E^+} + P_{E^\times} \over P_r} \approx {8 P_{\hodd} \over P_{\hat{r}}} \left({\kappa^2 z^2 \over a^2} \right) \approx 16 \epsilon
\ee
This fact, in conjuction with \eqref {eq:g_star_grav} and \eqref{gstarR}, leads to the prediction
\be\label{eq:consistency condition}
{g_{* \text{grav}} \over g_*} \approx {1 \over 64} {P_T \over P_S}.
\ee
The direction-dependent effects of a small persistent anisotropy during inflation on the tensor power spectrum are suppressed with respect to the direction-dependent effects on the scalar power spectrum by a number of order the tensor-to-scalar ratio.  This is a consistency condition for the model, given the constraint from observation, $\rh \ll \epsilon$.

\section{Conclusions}\label{sec:conclusions}
In this paper, we considered gauge-invariant perturbations in a class of models with a persistent background anisotropy.  After determining the quadratic action in terms of the dynamical fields, we computed the dominant direction-dependent effects of the background anisotropy on primordial power spectra. 

We showed that even a very small persistent anisotropy (with the anisotropy parameter much smaller than the slow-roll parameter $\epsilon$) can give rise to a dramatic direction-dependent effect on the primordial power spectra of dynamical fields. In an anisotropic background, the coupling between what reduce to the spin-1 and the spin-0 and spin-2 degrees of freedom in the isotropic case is extremely important.   We showed that such couplings give rise to the dominant direction-dependent contributions to the primordial power spectra of tensor and scalar perturbations.

There has been a fair amount of work on vector fields with time-dependent couplings that are put in by hand, assuming exponential expansion. We found that the amount of anisotropy in power spectra are quite sensitive to the details of how nonexponential the expansion is, and how long the expansion lasts. Perhaps this sensitivity is unsurprising in light of the no-hair theorem.

We found that for a given scale $| \vec{k}|$, the curvature power, $P(\vec{k})$, is minimized when $\vec{k}$ points along the preferred direction.\footnote{In other words, we found that $g_*$ is negative.} We attribute this feature to the fact that, in the class of models we considered, the preferred direction is expanding more slowly than other directions.

We showed that anisotropic effects are more pronounced in the scalar power spectrum than in the tensor power spectra.  In fact, we showed that the direction-dependent effects on the tensor power spectrum are suppressed with respect to the direction-dependent effects on the scalar power spectrum by a number of order the tensor-to-scalar ratio. \emph{A priori} one might have expected that the tensor power spectra and the scalar power spectrum would develop fractional direction dependence of the same magnitude.  We find that this is not the case.  

Finally, upon examination of the quadratic action for all dynamical degrees of freedom, we find no indication of instabilities in this model.  This should not be surprising since the matter stress-energy satisfies the dominant energy condition.  

We did not calculate the cross correlation between tensor and scalar perturbations. But one can see from the form of the quadratic action\footnote{See equations \eqref {eq:M1A} - \eqref{eq:evenAction}.} that such a nonzero, direction-dependent correlation should exist. The cross-correlation effect will be small compared to the direction-dependent effect on the curvature power spectrum, but it could be interesting.


\acknowledgements
We thank Sean Carroll, Mark Wise, and Matthew Johnson for their helpful comments.

\appendix

\section{Parametrization of perturbations}\label{app:parameterization}

In the following we use many of the same conventions and notation as in \cite{Pereira:2007yy}.  Since the background space-time is homogeneous, we decompose our perturbations into Fourier modes
\be
\delta(x^i,\eta) = \int {d^3 k \over (2\pi)^{3}}~ e^{ i k_j x^j} \delta (k_i, \eta).
\ee

For a given Fourier mode, characterized by the time-independent wave vector $k_i$, we form an orthonormal basis $\{e^1_i,e^2_i\}$ for the subspace perpendicular to the wave vector such that 
\be
\g^{ij} e^a_i e^b_j = \d^{ab} ~~~ \text{and} ~~~ \g^{ij} e^a_i k_j = 0.
\ee
Here $\gamma_{ij}$ is the spatial metric defined in \eqref{Bianchi1}.  Such an orthonormal basis for the spatial hypersurfaces is uniquely defined up to a spatial rotation about the wave vector $k_i$.  To remain properly normalized with the above normalization condition, these basis vectors must be time-dependent.

For definiteness, and without loss of generality, we will take wave vectors to be of the form $k_i = (k_1, k_2, 0)$.  The basis vectors can then be written as
\begin{equation}\label{eq:basis}
e^1_i = \left( - {e^{-3\beta} k_2 \over \sqrt{k^2}},~ {e^{3\beta} k_1 \over \sqrt{k^2}}, ~0\right) ~~\text{and}~~ e^2_j = \left(0,~ 0, ~e^{\beta}\right),
\end{equation}
where $\g^{ij} k_i k_j = k^2$.

It turns out that there always exists a choice of basis vectors $e^1_i$ and $e^2_j$ that results in the basis vectors having definite sign under what we will call 
\begin{equation}\label{k-parity}
\text{k parity:}~~~ k_i \rightarrow -k_i.
\end{equation}  
Our basis \eqref{eq:basis} is such that under k parity, $e^a_i \rightarrow (-1)^a e^a_i$.  Such a choice of basis is now unique up to discrete spatial rotations around the $k_i$ axis by multiples of $\pi/2$.  

We parametrize the most general perturbations about the background Bianchi I metric \eqref{Bianchi1} in the standard way,
\be
ds^2 = -a(\eta)^2\left[(1+2A) d\eta^2 + 2 B_i dx^i dt +  (\g_{ij}(\eta) + h_{ij}) dx^i dx^j \right].
\ee
Following \cite{Pereira:2007yy},
\begin{align}
 B_i &= \partial_i B + \bar{B}_i \\
h_{ij} &= 2 C \left(\g_{ij} + {\s_{ij} \over \cal{H}}\right) + 2 \partial_i \partial_j E + 2 \partial_{(i} E_{j)} + 2 E_{ij}
\end{align}
where $\s_{ij} = {1 \over 2} \g'_{ij}$ and $\cal{H} = \alpha'$ and also,
\begin{align}
\g^{ij} \partial_i \bar{B}_j = 0,~~~ \g^{ij} \partial_i E_j = 0,~~~\g^{ij} \partial_i E_{jk} = 0 ~~~ \text{and}~~~\g^{ij} E_{ij} = 0.
\end{align}
We parametrize perturbations of the inflaton field and the electromagnetic field by $\d \phi$ and $\d F_{\mu \nu}$, respectively.

One can show that the following are $U(1)$ gauge and diffeomorphism invariant variables,
\begin{align}
\Phi(k) &= A + {1 \over a(\eta)} \left( a \left(B - {(k^2 E)' \over k^2}\right)\right)', \\
\Psi(k) &= - C - {a'(\eta) \over a(\eta)} \left[ B -  {(k^2 E)' \over k^2}\right], \\
\Phi^i(k) &= \bar{B}^i - (E^i)', \\
E_{ij},\\
\chi(k)&= \d \phi + \phi'(\eta) \left[ B - {(k^2 E)' \over k^2}\right], \\
\Phi^F_{ij}(k) &= \d F_{ij} + 2 \bar{F}_{\eta[i} i k_{j]} \left[B - {(k^2 E)' \over k^2}\right], \\
\Phi^F_i(k) &= \d F_{\eta i} - \g^{jk} \bar{F}_{\eta j} i k_i (i k_k E + E_k) + \left(\bar{F}_{\eta i}\left[ B -  {(k^2 E)' \over k^2}\right] \right)'.
\end{align}

The perturbation in the gauge field can be decomposed along directions transverse and parallel to the spatial wave vector:
\begin{equation}\label{eq:emgaugefields}
\delta A_i =  (i \delta A^{(\perp, +)}(k,\eta))e^1_i + (\delta A^{(\perp, -)}(k,\eta) )e^2_i +(i \delta A^{\parallel}(k,\eta)))\hat{k}_i,
\end{equation}
where the amplitudes $\delta A^{(\perp, \pm)}(k,\eta)$ are $U(1)$ gauge invariant.\footnote{The factors of $i$ accompanying some perturbations is to ensure that the relation $\delta^*(k,\eta) = \delta(-k, \eta)$ holds for all Fourier amplitudes.} In $A_0 = E = B = B_i = 0$ gauge the electromagnetic gauge fields $\delta A^{(\perp, \pm)}(k,\eta)$ are simply related to the gauge-invariant magnetic and electric field perturbations. In particular we may define
\begin{equation}
\delta A^{+}(k,\eta)  \equiv { i ({e^1})^i k^j\Phi^F_{i j} \over k^2}  \qquad \text{and}  \qquad \delta A^{-}(k,\eta) \equiv  -{({e^2})^i k^j\Phi^F_{i j} \over k^2} , 
\end{equation}
where $\g^{ij} k_i k_j = k^2$ and where spatial indices are understood to be raised and lowered with the spatial metric, $\gamma_{i j}$. The dynamical, gauge-invariant dynamical electromagnetic variables are $\delta A^{\pm}(k,\eta)$ as defined above and are equal to  $\delta A^{(\perp, \pm)}(k,\eta)$ as defined in \eqref{eq:emgaugefields} in $A_0 = E = B = B_i = 0$ gauge (a modified Newtonian gauge). 

The tensor perturbations, $E_{i j}$ are gauge-invariant by construction. We will further decompose the tensor perturbations by constructing the two independent symmetric traceless tensors that are transverse to the wave vector $k_i$.  We again follow \cite{Pereira:2007yy} and define these tensors as
\begin{align}
E_{ij} &= E^+ \e^+_{ij} + i E^\times \e^{\times}_{ij}, \\
\e^+_{ij} &= {e^1_i e^1_j - e^2_i e^2_j \over \sqrt{2}}, \\
\e^{\times}_{ij} &= {e^1_i e^2_j + e^2_i e^1_j \over \sqrt{2}}.
\end{align}
We have chosen this normalization since,
\be
\g^{ik} \g^{jl} \e^\lambda_{ij} \e^{\lambda'}_{kl} = \d^{\lambda \lambda'}.
\ee
Because we have chosen a basis with the property that, under k-parity, $e^a_i \rightarrow (-1)^a e^a_i$ these tensors have k-parity transformations $\e^+_{ij} \rightarrow + \e^+_{ij}$ and $\e^\times_{ij} \rightarrow - \e^\times_{ij}$.

We will take the Mukhanov-Sasaki scalar variable (which is conserved outside the horizon in the isotropic limit) to be
\be\label{eq:r}
r \equiv {\alpha' \over \bar{\phi}'} \chi + \Psi.
\ee
In a gauge with spatially flat slicing, this variable corresponds to minus the curvature perturbation, $- \zeta$, as defined, \emph{e.g.}, in \cite{Dodelson:2003ft}.  
 
Some of the variables listed are not dynamical and must be removed from the action using constraint equations.  There are a total of five dynamical variables in the theory.  In the isotropic limit, these variables correspond to two electromagnetic perturbations, two tensor perturbations and one scalar perturbation.  Furthermore, the action separates into uncoupled parts according to the transformation of fields under k parity: a piece including $E^+$, $\delta A^+$ and $r$ and one including $E^{\times}$ and $\delta A^-$.

\section{Quadratic action and Einstein's equations}\label{ap:einsteinHilbert}

Given a metric $g_{\m\n} = \bar{g}_{\m\n} + \delta g_{\m\n}$, the Einstein-Hilbert action to quadratic order in $\delta g_{\m\n}$ can be written as
\begin{eqnarray}
\d^{(2)} S_{EH} &=& \int d^4 x \sqrt{-\bar{g}} \Big\{{1 \over 4 \kappa^2} \bar{g}^{\m\n} (\bar{\nabla}^\alpha \delta g_{\beta \m})(\bar{\nabla}^\beta \delta g_{\alpha \nu}) - {1 \over 4 \kappa^2} \bar{g}^{\m\n} (\bar{\nabla}^\alpha \delta g_{\m\n})(\bar{\nabla}^\beta \delta g_{\alpha \beta}) \nonumber \\ &+& {1 \over 8 \kappa^2} \bar{g}^{\m\n} \bar{g}^{\r\s} (\bar{\nabla}^\alpha \delta g_{\m \n})(\bar{\nabla}_\alpha \delta g_{\r \s}) - {1 \over 8 \kappa^2} \bar{g}^{\m \n} \bar{g}^{\r\s} (\bar{\nabla}^\alpha \delta g_{\m \r})(\bar{\nabla}_\alpha \delta g_{\n \s}) \\
+ {1 \over 2 \kappa^2} \bar{R}^{\m\n} \bar{g}^{\r\s}(\d g_{\m\r})(\d g_{\n\s}) &-& {1 \over 4\kappa^2} \bar{R}^{\m\n} \bar{g}^{\r\s}(\d g_{\m\n})(\d g_{\r\s}) + {1 \over 8\kappa^2}\bar{R} \,(\bar{g}^{\m\n} \d g_{\m\n})^2  -{1 \over 8\kappa^2} \bar{R} \,\bar{g}^{\m\n}\bar{g}^{\r\s} (\d g_{\m\r} \d g_{\n\s})  \Big\} \nonumber
\end{eqnarray}
after dropping boundary terms.   In the above equation, the covariant derivatives ($\bar{\nabla}$) are compatible with the background metric
\begin{equation}
\bar{\nabla}_\alpha \bar{g}_{\m\n} = 0.
\end{equation}

We used this form of the action and our parameterization to compute Einstein's equations.  In particular, the first-order change in the components Einstein tensor can be written in the following way (in Newtonian gauge, where $E = B = B_i = 0$),
\begin{eqnarray} \label{deltaEtimetime}
a^2 \d G^\eta_\eta &=& -2 \Delta \Psi + 6 \ch \Psi' - \left({\Psi \over \ch}\right)' \s^2 + {\s^{ij} \over \ch} \partial_i \partial_j \Psi - \s^i_j \partial_i \Phi^j + (E^i_j)' \s_i^j + (6 \ch^2 - \s^2) \Phi - \half (\s^2)' {\Psi \over \ch} \\ \label{deltaEtimespace}
a^2 \d G^\eta_i &=& - \s^2 {\partial_i \Psi \over \ch} + \s_i^j \partial_j \left(\Phi + \left({\Psi \over \ch}\right)'\right) - 2 \partial_i(\Psi' + \ch \Phi) 
+ \half \D \g_{ij} \Phi^j \\ 
&-& 2 \s^j_k \partial_j E^k_i + \s_j^k \partial_i E^j_k + 3 \s_i^j\partial_j \Psi + {(\s_i^j)'\over \ch}\partial_j \Psi \nonumber \\ \label{deltaEspacespace}
a^2 \d G^i_j &=& \d^i_j \left[2 \Psi'' + (2 \ch^2 + 4 \ch') \Phi + \D(\Phi - \Psi) + 2 \ch \Phi' + 4\ch \Psi' \right] - \partial^i \partial_j(\Phi - \Psi) - 2 { \s^{(i}_k \over \ch} \partial _{j)}\partial^k \Psi  \\
&+& \s^i_j \left[ -\ch \left({\Psi' \over \ch^2}\right)' + \left({ \ch ' \over \ch^2}\right)' \Psi + {\D \Psi \over \ch} - \Phi'\right] + \d^i_j \left[ \s^2 (\Phi + (\Psi/\ch)') + {\s^{k l} \over \ch} \partial_k \partial_l \Psi \right]  \nonumber \\
&+& (E^i_j)'' - \D E^i_j + 2\ch (E^i_j)' - \s^l_k (E^k_l)' \d^i_j + \d^i_j (\s^k_l \partial_k \Phi^l) - 2 \ch \g^{ik} \partial_{(k} \Phi_{j)} \nonumber \\
&-& \g^{ik} \left[ \partial_{(k} \Phi'_{j)} - 2 \s^l_{(k} \partial_{|l|} \Phi_{j)}\right] + (\s^i_j)' \left[ 2 {\ch' \over \ch^2} \Psi - 2 {\Psi' \over \ch} - 2(\Phi + \Psi)\right] + \s^i_j \left[2 {\ch' \over \ch} \Psi - 4 \ch \Phi\right] \nonumber \\
&+& \half \d^i_j {{\s^2}' \over \ch} \Psi - {(\s^i_j)'' \over \ch} \Psi + 4 \ch \left[ \s^i_k E^k_j - \s_j^k E^i_k \right] + 2 \left[\s^i_k  E^k_j -\s_j^k E^i_k\right]' -5 \s^i_j \Psi' \nonumber \\
&+& 2 \ch \left[ \s^i_k \partial_j E^k - \s^k_j \partial_k E^i \right] + \left[(\s^i_k)' \partial_j E^k - (\s^k_j)' \partial_k E^i \right],\nonumber
\end{eqnarray}
where $'$ denotes derivatives with respect to conformal time and
\begin{equation}
{\cal H} = {a' \over a}, ~~~ \sigma_{ij} = {1\over 2} \gamma_{ij}'.
\end{equation}
In these equations, spatial indices are raised and lowered with $\gamma_{ij}$. 

Our expressions \eqref{deltaEtimetime} - \eqref{deltaEspacespace} match those of \cite {Pereira:2007yy} up to factors of the anisotropic stress, ${\sigma^i_j}'+ 2 \ch \sigma^i_j$, which in \cite {Pereira:2007yy} was set to zero. Note that the Einstein tensor is gauge covariant rather than gauge invariant.

\section{Diagonalizing a kinetic term}\label{ap:diagonal}

Suppose a kinetic term takes the form
\be
K = \half X^{\dagger}{'} X' + X^{\dagger}{'} M X+ X^{\dagger} M^{\dagger} X'
\ee
where $X$ is a vector of fields and $M$ is a time-dependent matrix.
Diagonalizing the kinetic term requires a change of variables
\be
X \longrightarrow V Y,
\ee
where $V$ is a time-dependent unitary matrix, such that 
\be
K \longrightarrow \half Y^{\dagger}{'} Y' + \text{total derivative} + Y^{\dagger} Q Y 
\ee
where $Q$ is some Hermitian matrix.  We can calculate directly that
\begin{equation}
K  = \half Y^{\dagger}{'} Y' + Y^{\dagger} \Big( V^{\dagger}{'} V +V^{\dagger}(M^{\dagger} - M) V \Big) Y'   + \text{total derivative} + Y^{\dagger} Q Y. \end{equation}
The kinetic term is diagonalized by a unitary matrix $V$ that satisfies
\be
V^{\dagger}{'} V =  - V^{\dagger}(M^{\dagger} - M) V \qquad \text{or equivalently} \qquad V V^{\dagger}{'}  =  M-M^{\dagger}  .
\ee

If $M$ were a time-independent matrix, then the kinetic term would be diagonalized by a constant unitary matrix $V$ such that 
\be
V^{\dagger} (M - M^{\dagger}) V = D
\ee where $D$ is a constant diagonal matrix.

\section{Estimates of integrals}\label{app:integrals}
In order to get a quantitative estimate of the effect of the anisotropic background on power spectra, we must estimate the integrals in \eqref{eq:Ph} - \eqref {eq:ChA}. We may take $\rh$, $\Sigma$, and the slow-roll parameters to be nearly constant. Then the relevant integrals are
\begin{equation}\label{eq:integralsA}
p^{-2} \int^\eta \sin\left( 2\psi_{\vec{p}}(\eta') - 2\psi_{\vec{p}}(\eta)\right) \, \alpha'(\eta')^2 \, \tilde{I}(p \eta',p \eta) d\eta', \qquad  p^{-2} \int^\eta \cos\left( 2\psi_{\vec{p}}(\eta') - 2\psi_{\vec{p}}(\eta)\right) \, \alpha'(\eta')^2 \, \tilde{I}(p \eta',p \eta) d\eta'
 \end{equation}
 \begin{equation}\label{eq:integralsB}
\qquad  \int^\eta (e^{2 n \beta(\eta')} - e^{2 n \beta(\eta_0)})  \, \tilde{I}(p \eta',p \eta) d\eta'  \qquad \text{and}  \qquad p^{-2} \int^\eta \alpha'(\eta')^2 \, \tilde{I}(p \eta',p \eta) d\eta'
\end{equation}
where $\tilde{I}(x,y)$ was defined in  \eqref{eq:Itilde} as
\begin{equation}
\tilde{I}(x,y) = \left(\frac{1}{2 x^2 y^2}-\frac{1}{2 x^2}+\frac{2}{x y}-\frac{1}{2 y^2}+\frac{1}{2}\right) \sin (2 x-2 y)+\left(\frac{1}{x^2 y}-\frac{1}{x y^2}+\frac{1}{x}-\frac{1}{y}\right) \cos(2 x-2 y).
\end{equation}

During slow-roll inflation, 
\be\label{eq:alphaprimeapprox}
\alpha'(\eta) = e^{\alpha(\eta)} H(\eta) \approx - {1 \over \eta}
\ee
\be
\psi_{\vec{p}}(\eta') - \psi_{\vec{p}}(\eta) \approx \left(\alpha(\eta') - \alpha(\eta)\right) {k_2 e^{-\beta_0} \over k_0} \sqrt{\rh}
\ee
\be
 (e^{2 n \beta(\eta')} - e^{2 n \beta(\eta_0)}) \approx 2 n \Sigma \left( \alpha(\eta') - \alpha(\eta_0) \right).
\ee

Let us define a new variable $z$ by\footnote{This is just a convenient dimensionless variable and is not equal to $a \phi'/\alpha'$ as in \eqref{eq:z}.}
\be
- p \eta  = e^{- z}.
\ee

From \eqref{eq:alphaprimeapprox} it's clear that 
\be
e^{z} \approx {a H \over p} \qquad \text{and so} \qquad z \approx \log(H/p) + \alpha.
\ee

We may thus rewrite the integrals \eqref {eq:integralsA} and \eqref {eq:integralsB} in terms of the variable $z$:

\begin{equation}\label{eq:integralsA2}
I_s \equiv p^{-1}  \int^{z_{*}} \sin \left( 2{\psi'_{\vec{p}} \over \alpha'} (z - z_{*}) \right) \, \tilde{I}(- e^{-z}, - e^{-z_{*}}) e^z dz, \qquad  I_c \equiv p^{-1} \int^{z_{*}} \cos\left( 2{\psi'_{\vec{p}} \over \alpha'} (z - z_{*}) \right) \, \tilde{I}(- e^{-z}, - e^{-z_{*}}) e^z dz
 \end{equation}
 \begin{equation}\label{eq:integralsB2}
I_1 \equiv p^{-1}  \int^{z_{*}} (z - z_0)\, \tilde{I}(- e^{-z}, - e^{-z_{*}}) e^{-z} dz, \qquad  I_2 \equiv p^{-1} \int^{z_{*}}\tilde{I}(- e^{-z}, - e^{-z_{*}}) e^z dz
\end{equation}
where $z_*$ is the value of $z$ at the end of inflation and
 \be {\psi'_{\vec{p}} \over \alpha'}  \equiv  {p_2 e^{-\beta_0} \over p} \sqrt{ \rh} . \ee

The function 
\be
 \tilde{I}(- e^{-z}, - e^{-z_{*}}) e^z
\ee
oscillates rapidly with growing amplitude for $z < 0$. See Fig. \ref{fig:integrand}. For $z > 0$ and values of $z_*$ on the order of tens, the function is well approximated by a constant 
\be
 \tilde{I}(- e^{-z}, - e^{-z_{*}}) e^z \approx - {2 \over 3} e^{2 z_*} \qquad 0 < z < z_*.
\ee
The constant can be found by expanding the function about $z_* = \infty $ and then about  $z = \infty$.

{
\begin{figure}
 \centering
\includegraphics[width=0.5\textwidth]{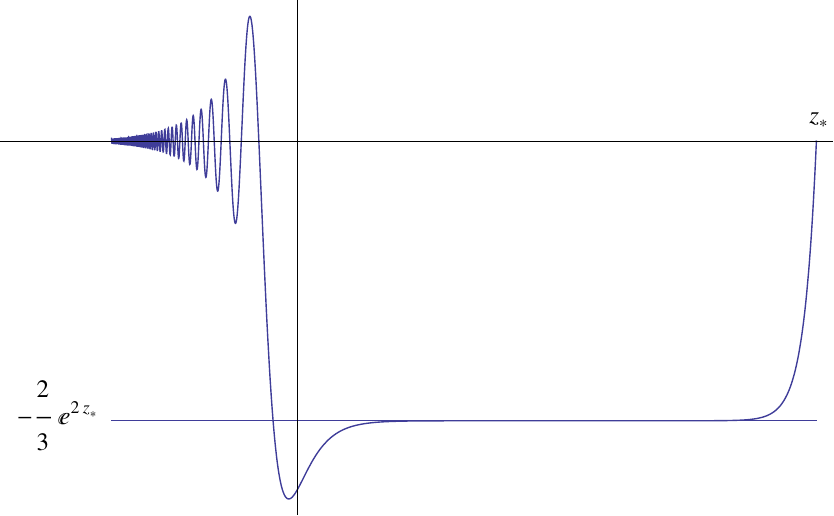}
\caption{The function $e^z \tilde{I}(-e^z,-e^{-z_*})$ on a linear scale. The axes cross at the point $\{ 0,0 \}$. For $0 < z < z_*$ the function is well approximated by $- {2 \over 3} e^{2 z_*}$. The frequency of oscillation for $z < 0$ does not vary much as $z_*$ increases---only the amplitude changes. The plot above was generated using $z_* = 15$.}\label {fig:integrand}
\end{figure}
}

The contribution of terms that go like $I_1$ will be subdominant compared to contributions from terms proportional to the other integrals\footnote{The contribution from $I_1$ can be important if inflation lasts a very long time --- on the order of $10^3$ $e$-folds.}, so we will not bother to calculate $I_1$. Since the dominant contribution to the other integrals will occur when $z > 0$ (which corresponds to after horizon crossing) we may approximate the integrals by
\begin{equation}
I_s \approx - {2 \over 3} e^{2 z_*} p^{-1}  \int^{z_{*}}_0 \sin \left(2 {\psi'_{\vec{p}} \over \alpha'} (z - z_{*}) \right)  dz = - {2 \over 3} e^{2 z_*} p^{-1} \left({2\psi'_{\vec{p}} \over \alpha'}\right)^{-1} \left( \cos \left({2\psi'_{\vec{p}} \over \alpha'} z_*\right) - 1 \right), 
\end{equation}
\begin{equation} 
I_c \approx - {2 \over 3} e^{2 z_*} p^{-1}  \int^{z_{*}}_0 \cos \left(2 {\psi'_{\vec{p}} \over \alpha'} (z - z_{*}) \right)  dz = - {2 \over 3} e^{2 z_*} p^{-1} \left({2\psi'_{\vec{p}} \over \alpha'}\right)^{-1} \left(  - \sin\left( {2\psi'_{\vec{p}} \over \alpha'} z_*\right)\right)
 \end{equation}
\begin{equation}
 I_2 = p^{-1} \int^{z_{*}}\tilde{I}(- e^{-z}, - e^{-z_{*}}) e^z dz \approx - {2 \over 3} e^{2 z_*} p^{-1} z_*.
\end{equation}
Modes of astrophysical interest crossed the horizon about $60$ $e$-folds---plus or minus a few---before the end of inflation. Such modes of astrophysical interest therefore correspond to $z_* \approx 60$. 

\newpage

\bibliography{hairy-bib-2}

\end{document}